 \documentclass[final,3p,times,twocolumn]{elsarticle}

\usepackage{amssymb}

\usepackage[usenames,dvipsnames]{color}
\usepackage{ulem} 
\usepackage[
colorlinks=true, citecolor=blue, urlcolor=blue, linkcolor=blue,
setpagesize=false
]{hyperref}
\usepackage{gensymb}

\journal{Current Opinion in Solid state and Materials Science}

\begin{document}

\begin{frontmatter}



\title{Recent advances in MXenes: from fundamentals to applications}


\author[label1]{Mohammad Khazaei\corref{cor1}}
\ead{khazaei@riken.jp}
\cortext[cor1]{Corresponding author.}
\address[label1]{Computational Materials Science Research Team, RIKEN Center for Computational Science (R-CCS), Kobe, Hyogo 650-0047, Japan}
\author[label2]{Avanish Mishra}
\address[label2]{Materials Research Centre, Indian Institute of Science, Bangalore 560012, India}
\author[label3]{Natarajan S. Venkataramanan}
\address[label3]{Department of Chemistry, School of Chemical and Biotechnology, SASTRA University, Thanjavur, 613 401, India}
\author[label2]{Abhishek K. Singh}
\author[label1,label4,label5]{Seiji Yunoki}
\address[label4]{Computational Condensed Matter Physics Laboratory, RIKEN Cluster for Pioneering Research (CPR), Wako, Saitama 351-0198, Japan}
\address[label5]{Computational Quantum Matter Research Team, RIKEN Center for Emergent Matter Science (CEMS), Wako, Saitama 351-0198, Japan}

\begin{abstract}
The family of MAX phases and their derivative MXenes are continuously growing in terms of both crystalline and composition varieties. In the last couple of years, several breakthroughs have been achieved that boosted the synthesis of novel MAX phases with ordered double transition metals and, consequently, the synthesis of novel MXenes with a higher chemical diversity and structural complexity, rarely seen in other families of two-dimensional (2D) materials. Considering the various elemental composition possibilities, surface functional tunability, various magnetic orders, and large spin$-$orbit coupling, MXenes can truly be considered as multifunctional materials that can be used to realize highly correlated phenomena. In addition, owing to their large surface area, hydrophilicity, adsorption ability, and high surface reactivity, MXenes have attracted attention for many applications, e.g., catalysts, ion batteries, gas storage media, and sensors. Given the fast progress of MXene-based science and technology, it is timely to update our current knowledge on various properties and possible applications. Since many theoretical predictions remain to be experimentally proven, here we mainly emphasize the physics and chemistry that can be observed in MXenes and discuss how these properties can be 
tuned or used for different applications. 
\end{abstract}


\end{frontmatter}







\section{INTRODUCTION}

Owing to their unique electronic structures and large surface areas, two-dimensional (2D) materials are promising candidates for many electronic and energy applications. Hence, 
the synthesis, properties, and applications of novel 2D materials have currently become one of the most exciting areas of interest in science and technology. Single layers of graphene, boron nitride (BN), transition-metal dichalcogenides (MoS$_2$, WS$_2$, $, etc.$), and phosphorene have been successfully obtained from their bulk van der Waals layered structures~\cite{K.S.Novoselov2016}. Recently, it has been shown that by using a combination of chemical exfoliation and sonication, the synthesis and mass production of 2D materials from three-dimensional (3D) layered compounds with chemical bonding between the layers are also feasible~\cite{M.Naguib2011,M.Naguib2012}.
In this regard, it has been demonstrated that using hydrofluoric acid (HF) solutions and sonication, some members of the MAX phase family~\cite{M.W.Barsoum2000} can be exfoliated into 2D transition-metal carbide and nitride layers, so-called MXenes~\cite{M.Naguib2011,M.Naguib2012}. During exfoliation, depending on the type of chemical environment,
a mixture of F, O, or OH groups terminates the surface of MXenes~\cite{M.A.Hope2016,J.Halim2016}. 
It should be noted that HF is not the only etchant but  
LiF+HCl~\cite{M.Ghidiu2014_2} and NH$_4$HF$_2$~\cite{J.Halim2014} have also been used for exfoliation.

\begin{figure}[t]
\centering
\includegraphics[width=1.0\columnwidth]{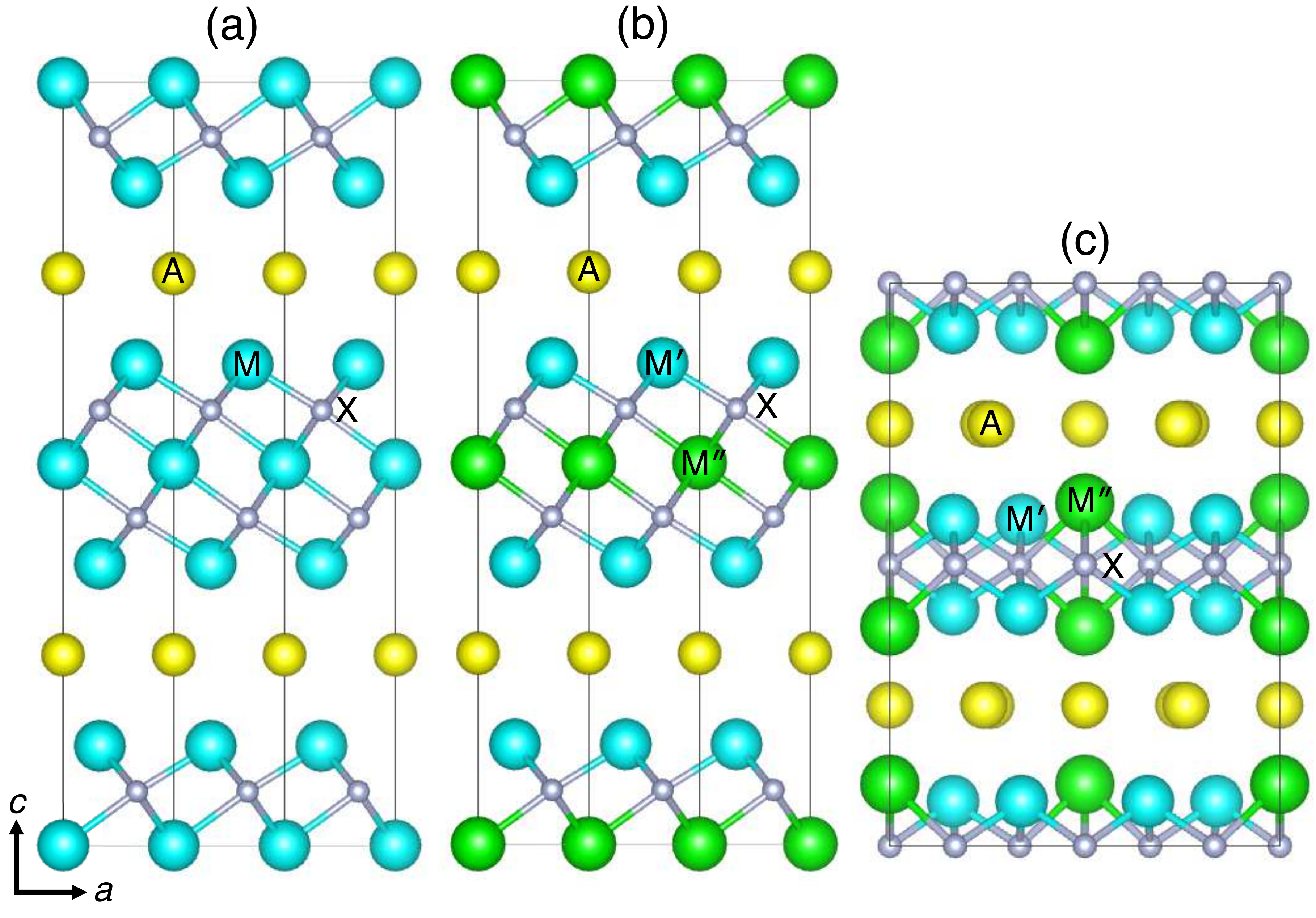}
  \caption{Crystal structure of (a) M$_3$AX$_2$, (b) out-of-plane ordered double transition-metal M$'_2$M$''$AX$_2$, 
  and (c) in-plane double transition-metal (M$'_{2/3}$M$''_{1/3}$)$_2$AX. Solid lines indicate the border of the unit cells. 
}
  \label{fig:MAXPhase}
\end{figure}

The applications of solid MAX phases are somehow limited to their metallic, high-temperature stabilities, and superior mechanical properties~\cite{M.W.Barsoum2011,M.Khazaei2014_1}. However, owing to their vast compositional possibilities, MAX phases include many crystalline and numerous alloy members. When these MAX phases are exfoliated into 2D MXenes, depending on the transition metals at the surface, thickness, and surface functionality, they exhibit quite different electronic, magnetic, optical, and electrochemical properties that are rarely seen in their original MAX phases. There are many theoretical studies in the literature that predict MXenes possess excellent electronic properties and many device applications~\cite{M.Khazaei2017}. 
However, most of these predictions have not been realized yet. This is because many of these predictions are based on the ideal crystal structure of MXenes without any M and X atomic vacancies and with a homogeneous surface termination having one type of the F, O, or OH group. Unfortunately, the formation of such ideal MXenes is still an experimental challenge. Recently, there are some theoretical attempts to investigate the properties of MXenes with a mixture of surface functional groups~\cite{N.M.Caffrey2018,T.Hu2018}, and conversely there have been some experimental attempts to synthesize MXenes with a uniform type of attached chemical group by using temperature control~\cite{I.Persson2018}. Most experimental investigations have mainly focused on the excellent electrochemical behavior of MXenes for energy storage as ion batteries, gas storage, and various catalysis applications because of their large exposed surface area, hydrophilic nature, adsorption ability, and surface activities. 
The latest applications and developments of MXenes have been summarized in the recent review 
article~\cite{M.Khazaei2017,B.Anasori2017_1,B.M.Jun2018,J.Pang2019,A.L.Ivanovskii2013,N.K.Chaudhari2017,J.Zhu2017,H.Wang2018,X.Li2018,X.Zhang2018,Y.Zhang2018,K.Hantanasirisakul2018,H.Lin2018}.

Although most theoretical studies are conducted using ideal MXenes and are mostly performed on the basis of density functional theory (DFT), 
which is known to have some drawbacks related to the accurate prediction of the band gaps and van der Waals or strongly correlated interactions, the information provided by those studies is
still valuable for understanding the physics and chemistry of MXenes. 
In this review, we first provide a solid background of the structural properties of MAX phases and 
MXenes, and give insights into the possibility of the exfoliation of MAX phases into 2D MXenes. 
We then summarize the electronic, magnetic, and optical properties of MXenes predicted from the theoretical studies. 
Finally, we discuss the manifestation of these properties in various applications such as the photocatalysis, electrocatalysis, 
chemical catalysis, ion batteries, gas storage, Schottky barriers, thermo-, ferro-, and piezoelectric applications. 
This review can be considered as complementary to and an update of our previous review on the electronic properties and applications of MXenes~\cite{M.Khazaei2017}.

\section{Crystalline MAX Phases and their 2D derivative MXenes}

\begin{figure}[t]
\centering
\includegraphics[width=1.0\columnwidth]{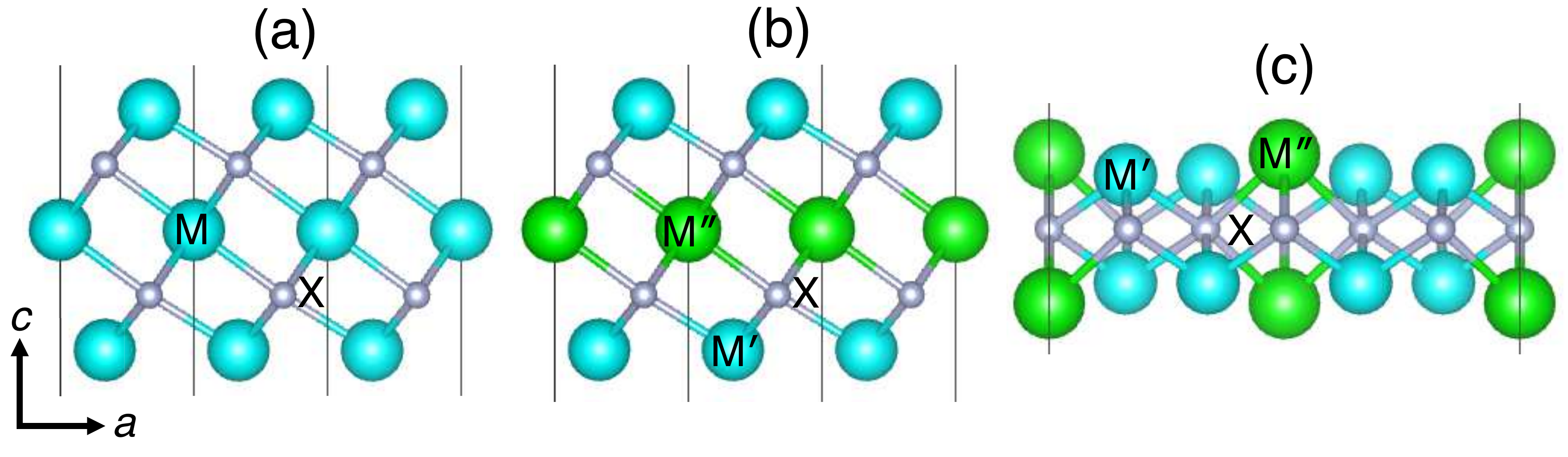}
  \caption{Crystal structure of (a) M$_3$X$_2$, (b) out-of-plane ordered double transition-metal M$'_2$M$''$X$_2$, 
  and (c) in-plane double transition-metal (M$'_{2/3}$M$''_{1/3}$)$_2$X MXenes. Solid lines indicate the border of the unit cells. 
}
  \label{fig:MXene}
\end{figure}

MAX phases are a family of solids with layered hexagonal structures and a space group symmetry of P6$_3$/mmc (No. 194), whose chemical compositions are traditionally known by the chemical formula M$_{n+1}$AX$_n$, where $n$ = 1, 2, or 3, ``M" is an early transition metal (Sc, Ti, Zr, Hf, V, Nb, Ta, Cr, or Mo), ``A" is an element from groups III$-$VI in the periodic table (Al, Ga, In, Tl, Si, Ge, Sn, Pb, P, As, Bi, S, or Te), and ``X" is carbon and/or nitrogen~\cite{M.W.Barsoum2000}. As seen in Fig.~\ref{fig:MAXPhase}(a), each layer of X atoms is sandwiched between two layers of transition metals and every two M$_{n+1}$X$_n$ layers are interleaved with a layer of A atoms.  During an acid treatment, the ``A" atoms are washed from the MAX phase structure, resulting in multiple layers of M$_{n+1}$X$_n$, which can be dispersed into monolayers or a few layers by using sonication. 
As examples, the crystal structures of M$_3$AX$_2$ and M$_3$X$_2$ are shown in Figs.~\ref{fig:MAXPhase}(a) and~\ref{fig:MXene}(b), respectively. Inherited from the MAX phases, the 2D M$_{n+1}$X$_n$ layers also have hexagonal symmetry and in analogy to graphene, which has hexagonal symmetry, they have been named MXenes. Over 70 different M$_{n+1}$AX$_n$
MAX phases exist experimentally. Among them, Ti$_2$AlC, Ti$_2$AlN, V$_2$AlC, Nb$_2$AlC, Ti$_3$AlC$_2$, Ti$_3$SiC$_2$, Zr$_3$AlC$_2$, 
Ti$_4$AlN$_3$, V$_4$AlC$_3$, Nb$_4$AlC$_3$, and Ta$_4$AlC$_3$ have 
already been exfoliated into Ti$_2$C, Ti$_2$N, V$_2$C, Nb$_2$C, Ti$_3$C$_2$, Zr$_3$C$_2$, 
Ti$_4$N$_3$, V$_4$C$_3$, Nb$_4$C$_3$, and Ta$_4$C$_3$ MXenes~\cite{M.Naguib2012,M.Naguib2013,M.Ghidiu2014,J.Zhou2016,P.Urbankowski2016,B.Soundiraraju2017,M.Alhabeb2018,M.H.Tran2018}. 

\begin{figure*}[t]
\centering
\includegraphics[scale=0.76]{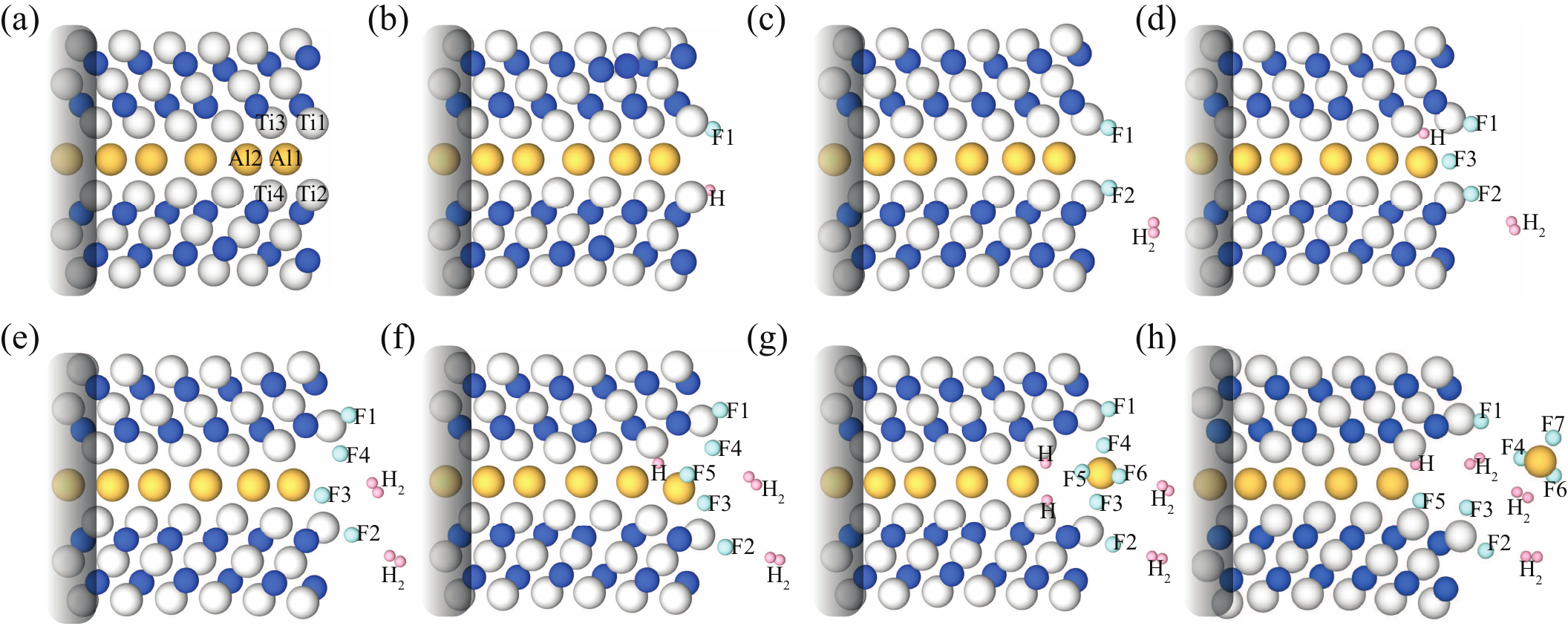}
  \caption{Schematic atomic configurations for chemical exfoliation of Ti$_3$AlC$_2$ using HF acid, where AlF$_3$ is extracted from the crystal: (a) pristine Ti$_3$AlC$_2$ and (b) one, (c) two, (d) three, (e) four, (f) five, (g) six, 
  and (h) seven HF molecules intercalated into Ti$_3$AlC$_2$~\cite{P.Srivastava2016}. 
  White, blue, orange, cyan, and pink spheres represent Ti, C, Al, F, and H atoms, respectively.  
}
  \label{fig:chemexfoliation}
\end{figure*}

There are many experimental observations that a solid solution of different transition metals (M$'$ and M$''$) and ``A" elements  (A$'$ and A$''$) or an``X" element (C and N) results in the formation of alloy MAX phases and consequently the synthesis of various alloy MXenes. In this regard, 
alloy TiNbC, (Ti$_{0.5}$Nb$_{0.5}$)$_2$C, (V$_{0.5}$Cr$_{0.5}$)$_3$C$_2$, Ti$_3$CN, 
(Nb$_{0.8}$Ti$_{0.2}$)$_4$C$_3$, and  
(Nb$_{0.8}$Zr$_{0.2}$)$_4$C$_3$ MXenes have already been experimentally fabricated~\cite{M.Naguib2012,J.Yang2016_2}. 
However, it has recently been shown that by the appropriate selection and stoichiometry of M$'$ and M$''$, A, and X, it is possible to synthesize two types of novel MAX phases with ordered double transition metals, named oMAX and iMAX phases. oMAX phases are MAX phases with 
out-of-plane ordered double transition metals, M$'_2$M$''$AX$_2$ and M$'_2$M$''_2$AX$_3$. Figures~\ref{fig:MAXPhase}(b) and~\ref{fig:MXene}(b) show the 
crystal structure of M$'_2$M$''$X$_2$ and its corresponding 2D M$'_2$M$''$X$_2$ MXene, respectively ~\cite{B.Anasori2015_1,B.Anasori2015_2,R.Meshkian2017}. Similar to the structures of traditional M$_{n+1}$AX$_n$ MAX phases, the layers of transition metals in oMAX phases are occupied with only one type of M or M$'$ and retain the same crystallographic space group symmetry of P6$_3$/mmc. Interestingly, their corresponding M$'_2$M$''$X$_2$ and M$'_2$M$''_2$X$_3$ MXenes have unique configurations, as seen in Fig.~\ref{fig:MXene}(b), such that the outer layer of the MXene consists of a transition metal M$'$ only and the inside layers consists of a transition metal M$''$ only. Mo$_2$TiAlC$_2$, Mo$_2$Ti$_2$AlC$_3$, Cr$_2$TiAlC$_2$, and Mo$_2$ScAlC$_2$ have already been synthesized and exfoliated into their corresponding 2D Mo$_2$TiC$_2$, Mo$_2$Ti$_2$C$_3$, Cr$_2$TiC$_2$, and Mo$_2$ScC$_2$ MXenes, respectively~\cite{B.Anasori2015_1,B.Anasori2015_2,R.Meshkian2017}.

iMAX phases are MAX phases with in-plane ordered double transition metals, (M$'_{2/3}$M$''_{1/3}$)$_2$AX. Different from traditional MAX and oMAX phases, each layer of transition metals 
in iMAX phases contains both M$'$ and M$''$
[see Fig.~\ref{fig:MAXPhase}(c)]~\cite{Q.Tao2017,M.Dahlqvist2017,J. Lu2017,R.Meshkian2018,L.Chen2018,M.Dahlqvist2018,J.Halim2018}. In contrast to previous MAX phases that consist of A atoms forming a hexagonal lattice, A atoms are located on a Kagome-like lattice in iMAX phases. This occurs because of the different atomic sizes in which M$'$ $<$ M$''$ such that the M$''$ atoms cause the A atoms to deviate from the hexagonal lattice to the Kagome-like lattice~\cite{M.Dahlqvist2018}. 
The space group symmetry of iMAX phases is $C2/c$.  
Accordingly, after exfoliation, the 2D MXenes obtained from iMAX phases include two 
different types of transition metals in each layer of transition metals [see Fig.~\ref{fig:MXene}(c)]. 
In analogy to 2D MXenes, the 2D structures derived from iMAX phases can 
accordingly be called as iMXenes. The iMAX phases of (M$'_{2/3}$M$''_{1/3}$)$_2$AlC (M$'$ = Cr, Mo, W; M$''$ = Sc, Y),
and (M$'_{2/3}$Zr$_{1/3}$)$_2$AlC (M$'$ = V, Cr) have already been experimentally fabricated and most of them 
have already been exfoliated. 
 However, during exfoliation process, Sc/Y and Al are dissolved, resulting in the formation of 
2D M$'_{1.33}$C MXenes with ordered vacancies~\cite{Q.Tao2017,M.Dahlqvist2017,J. Lu2017,R.Meshkian2018,L.Chen2018,M.Dahlqvist2018}. 
It is noteworthy that (Nb$_{2/3}$Sc$_{1/3}$)$_2$AlC has also been fabricated in the solid solution form 
and its exfoliation results in 2D Nb$_{1.33}$C MXene with randomly distributed vacancies~\cite{J.Halim2018}.  



\section{Exfoliation of MAX phases to MXenes}
The chemical exfoliation of MAX phases is a complex process with many reaction kinetics and dynamics, which makes it difficult to model or simulate them in detail.
Nevertheless, we can still obtain some valuable information about the reaction processes 
using molecular dynamic simulations~\cite{P.Srivastava2016}
or find some good candidates that can likely be exfoliated into 2D MXenes by examining the bond strengths 
and exfoliation energies through static calculations~\cite{M.Khazaei2018_1}.

\subsection{Dynamics perception}

The exfoliation process of MAX phases into 2D MXenes necessarily involves the elimination of the A element using an acid treatment, e.g., a HF solution. 
The exfoliation of Ti$_3$AlC$_2$ into 2D Ti$_3$C$_2$ has been 
theoretically studied in the presence of water and HF using \textit{ab initio} molecular dynamics 
simulations~\cite{P.Srivastava2016}. It was shown 
that after the dissociation of HF molecules into H and F radicals, they are adsorbed
at the edge Ti atoms, which weakens the Al$-$Ti bonds, subsequently opening an interlayer gap. 
The interlayer gap, in turn, facilitates the further penetration of HF molecules. This leads to the formation of 
AlF$_3$ and H$_2$, which are eventually extracted from
the MAX phase. After these processes, the fluorinated MXene is left 
behind~\cite{P.Srivastava2016}. This process is illustrated in Fig~\ref{fig:chemexfoliation}.

 \begin{figure}[t]
\centering
  \includegraphics[width=1.0\columnwidth]{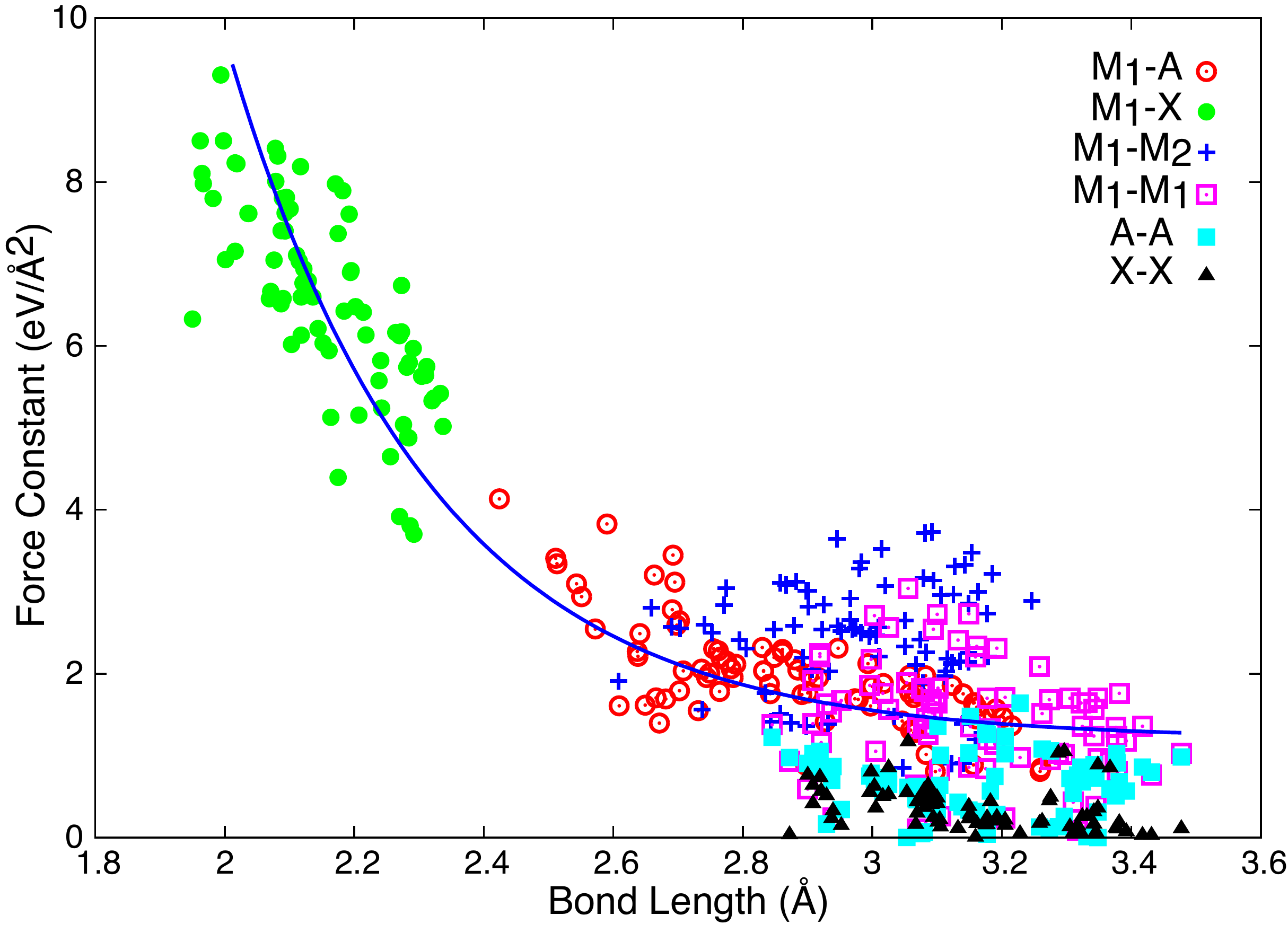}
  \caption{Force constant vs. bond length of different bonds for 
  82 different MAX and oMAX phases~\cite{M.Khazaei2018_1}. The line indicates the general trend.}
  \label{fig:bonddistance}
\end{figure}

\subsection{Static perception}

It is time-consuming and complicated to address the question of which MAX phases can likely be exfoliated into 2D MXenes
by using molecular dynamics simulations. However, static calculations are also helpful to screen the bond properties of many MAX phase structures in a short time.

Theoretically, on the basis of tensile and shear module analyses, 
the mechanical exfoliation of M$_2$AlC (M = Ti, Zr, Hf, V, Nb, Ta, Cr, Mo, and W) MAX phases into 2D M$_2$C MXenes has been investigated~\cite{Z.Guo2015}. 
It was shown that by applying large tensile stress, the M$-$Al bonds are broken, 
leading to the separation of M$_2$C and Al layers~\cite{Z.Guo2015}. 
Although MXenes are made by the chemical exfoliation of MAX phases, their synthesis through the mechanical exfoliation of MAX phases may also be possible.
This is because in some of the MAX phases, their elastic constants C$_{11}$ in the $ab$ plane are larger than 
C$_{33}$ perpendicular to the plane, which indicates that the overall bonding in the $ab$ plane is stronger than that along the 
$c$ direction~\cite{M.Khazaei2014_1,M.Khazaei2014_2}. Therefore, if
C$_{33}$ is smaller than C$_{11}$, it might be feasible to break the M$-$A$-$M bonds under appropriate mechanical
tension without significantly damaging the M$-$X$-$M bonds \cite{M.Khazaei2014_1,M.Khazaei2018_1,M.Khazaei2014_2}.
This is an important theoretical observation because 
it reveals the possibility of the fabrication of MXenes by mechanical exfoliation of MAX phases~\cite{M.Yi2015}.
To the best of our knowledge, there are no experiments on the mechanical exfoliation of MAX phases. However, it would be a promising method to achieve the large-scale production of MXenes at an extremely low cost.

The evaluation of the bond strengths of the M$-$A and M$-$X bonds is a straightforward method to determine the favorable candidates 
for the exfoliation to 2D MXenes. MAX phases with strong M$-$X bonds and weak M$-$A bonds are the best candidates. 
For this purpose, the force constants of the M$-$A and M$-$X bonds and the exfoliation energies are useful quantities 
for revealing
 information about the bond strength 
and the ease of bond breaking during the exfoliation process~\cite{M.Khazaei2018_1}. As an example of such analyses, the force constants of
M$-$A, M$_1-$X, M$_1-$M$_1$, M$_1-$M$_2$, and X$-$X bonds 
are investigated in Fig.~\ref{fig:bonddistance}. 
Here, M$_1$ and M$_2$ represent transition metals placed on the first and second 
layers of transition metals adjacent to the A element. The X layer is located between the M$_1$ and M$_2$ layers. 
Results have been accumulated for 82 different MAX and oMAX phases~\cite{M.Khazaei2018_1}. 
The overall trend is that shorter bonds are stronger. Interestingly,
the force constants of the M$_1-$X bonds are significantly higher than those of the other bonds. 
 In other words, the M$_1-$X bonds are the strongest in 
MAX phases. The X$-$X and A$-$A bonds are the weakest among all bonds, as shown 
in Fig.~\ref{fig:bonddistance}~\cite{M.Khazaei2018_1}. 


Considering the static exfoliation energies of MAX phases into 2D MXenes for all MAX phases that have been  
already exfoliated experimentally, it is found that V$_2$AlC$_2$ has the largest exfoliation energy of 0.205 eV \AA$^{-2}$~\cite{M.Khazaei2018_1}. 
Therefore, it is expected that MAX phases with an exfoliation energy smaller than 0.205  eV \AA$^{-2}$ have a good chance to be exfoliated into MXenes. 
Based on the static exfoliation energy calculations and the force constant analyses, 
it was shown that over  37 crystalline MAX phases, as listed in Ref.~\cite{M.Khazaei2018_1}, can potentially be exfoliated into 2D MXenes. 
These results indicate that in the family of MAX phases, there exist some members whose M$-$A (M$-$X) bonds are sufficiently weak (strong) such that they can (cannot) be broken during 
 acid treatments. This explains why the exfoliation of some of the MAX phases into 2D MXenes
 is feasible experimentally.

\section{ Characteristics and emergent attributes of MXenes}

\subsection{{\color {blue} Geometry and energetics}}

The crystalline structures of pristine 2D MXene systems can be simply obtained by eliminating the ``A" element from the MAX phases. 
However, owing to the reactivities of transition metals, the surfaces of MXenes are usually terminated with a mixture of F, OH, and O.   
It is difficult to predict the electronic properties of MXenes using a model structure with the mixed adsorption of F, OH, and O on their surfaces.
Hence, the majority of theoretical studies have been carried out using models with the uniform adsorption of one of the chemical groups on the 
surfaces~\cite{M.Khazaei2017} and very few using mixture of these groups~\cite{N.M.Caffrey2018,T.Hu2018}. Although such ideal structures might be 
unrealistic at present, they help with the in-depth understanding of the physics and chemistry of MXenes and may guide and motivate ones 
to perform new experiments.

 \begin{figure}[t]
\centering
\includegraphics[width=1.0\columnwidth]{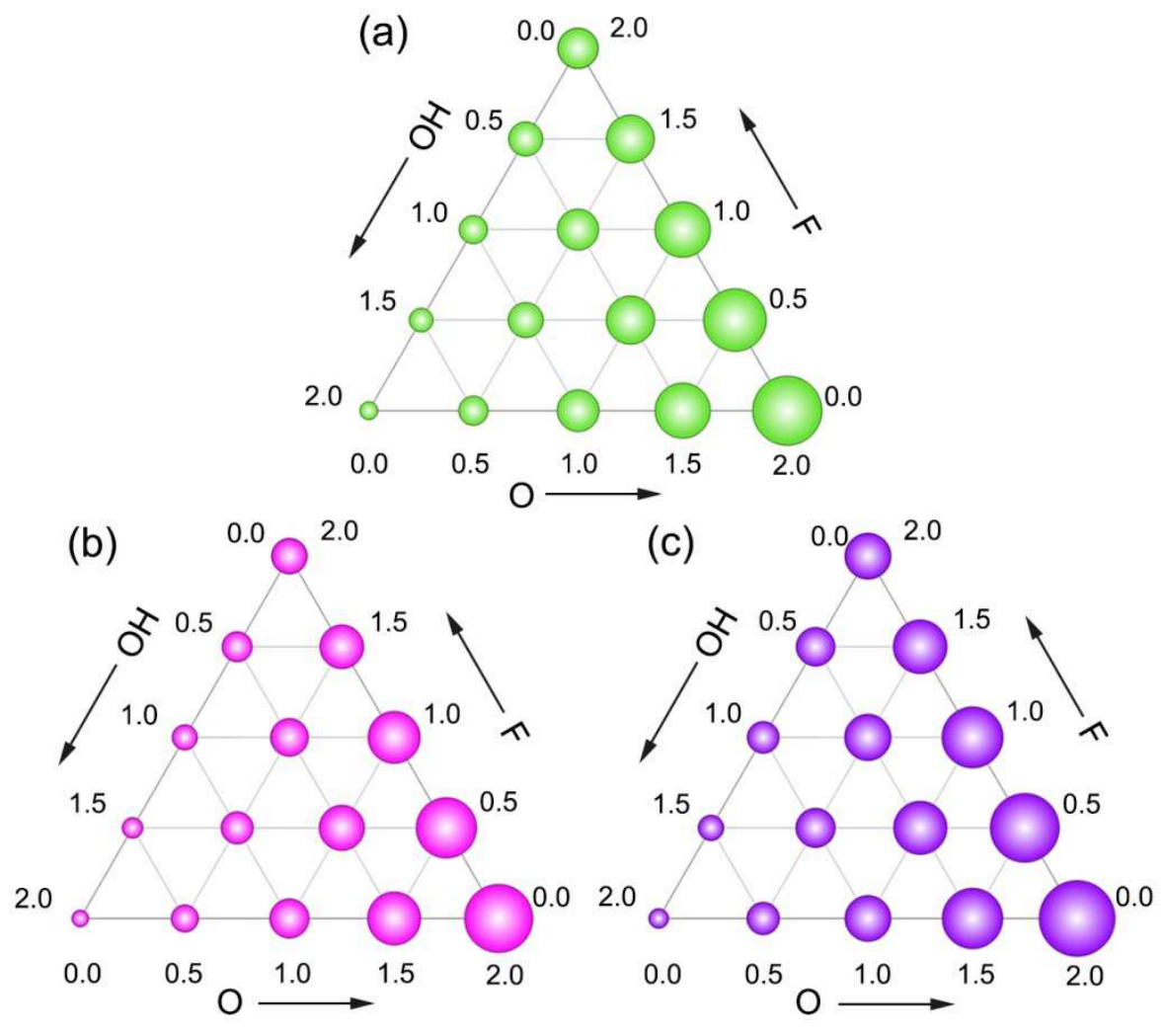}
  \caption{Stability of Ti$_3$C$_2$ (a), Ti$_2$C (b), and Nb$_4$C$_3$ (c) after functionalization with mixture of O$_x$F$_y$(OH)$_z$. 
  The sizes of spheres indicate the relative stabilities of the MXenes with various O$_x$F$_y$(OH)$_z$ functionalization~\cite{T.Hu2018}. 
}
  \label{fig:stability}
\end{figure}

On the surfaces of MXenes, there are two types of hollow sites; the sites at which there exists no X atom available under them 
and the sites at which there exists an X atom available under them. Therefore, according to the relative positions of the adsorbed termination groups at
the hollow sites, there are several choices for the uniform surface terminations of MXenes that should be taken into account for finding 
the lowest energy configurations before analyzing the electronic properties.  
It is noted that when F, OH, or O is adsorbed on top of 
transition metals on the surface, they usually move to hollow sites during the structural relaxation calculations. . 
Therefore, it is anticipated that the termination groups 
are favorably adsorbed at the hollow sites. As explained in the previous section, the M$-$X bonds 
are the stiffest bonds in MAX phases. Hence, the chemical groups (F, OH, or O) are adsorbed at hollow sites, 
where the bonding states of the M$-$X bond achieve their highest occupancy~\cite{A.Mishra2017}.
Further thermodynamic simulations find that when the surfaces of MXenes are fully functionalized,
they become thermodynamically more stable. This is understandable because, in various complexes and crystals, the number of ions surrounding a 
transition metal is often six, which results in the formation of a functionalized M$_2$XT$_2$ (T = F, OH, or O) MXene~\cite{M.Khazaei2013}. 
The dynamic stabilities of such fully functionalized MXenes have
already been proved by phonon calculations~\cite{U.Yorulmaz2016}.

Regarding the multilayer crystal structures of MXenes, similar to graphite, they may form in different stacking orders. Therefore, it is 
important to obtain the correct stacking order before examining the electronic structures of multilayer MXenes. 
The lowest energy coordinate structures of many of monolayer or multilayer 
MXenes functionalized with F, OH, or O can be found in the supporting information files of Refs.~\cite{M.Khazaei2013,M.Khazaei2014_3}.
These structures were determined based on the first-principles calculations using GGA/PBE method. 
However, it might be necessary to investigate the effect of van der Waals interactions in the stability of 
different stacking orders~\cite{Y.Xie2014_1}.

F, O, and OH form strong bonds with MXenes with large negative adsorption energies~\cite{M.Khazaei2013}.
 The different adsorption energies of F, OH, and O on MXenes can be explained by examining the effect of the octahedral crystal field, 
 which is created by carbon and the chemical groups, on the $d$ orbitals of the transition metal. 
 For example, a recent study on 
 Ti$_3$C$_2$T$_2$ (T= O, F, H, OH) has shown that owing to the generated crystal
 field, the 3$d$ orbitals of surface Ti atoms split to form pseudogaps, whose values predict stability in order 
 from highest to lowest of
 Ti$_3$C$_2$O$_2$, Ti$_3$C$_2$F$_2$, Ti$_3$C$_2$(OH)$_2$, Ti$_3$C$_2$H$_2$, and Ti$_3$C$_2$~\cite{T.Hu2017}. Moreover, the effect of 
mixture of F, OH, and O functionalization on the stability of Ti$_2$CO$_x$F$_y$(OH)$_z$, Ti$_3$C$_2$O$_x$F$_y$(OH)$_z$, and Nb$_4$C$_3$O$_x$F$_y$(OH)$_z$ has also been examined~\cite{T.Hu2018}. 
As shown in Fig.~\ref{fig:stability}, the stability found from these analyses follows the similar trend: the stability of these MXenes is enhanced from OH, F, to O~\cite{T.Hu2018}. In other words, fully O-terminated (OH-terminated) Ti$_2$C, Ti$_3$C$_2$, Nb$_4$C$_3$ are the most (least) energetically favorable MXenes.

 \begin{figure}[t]
\centering
\includegraphics[width=1.0\columnwidth]{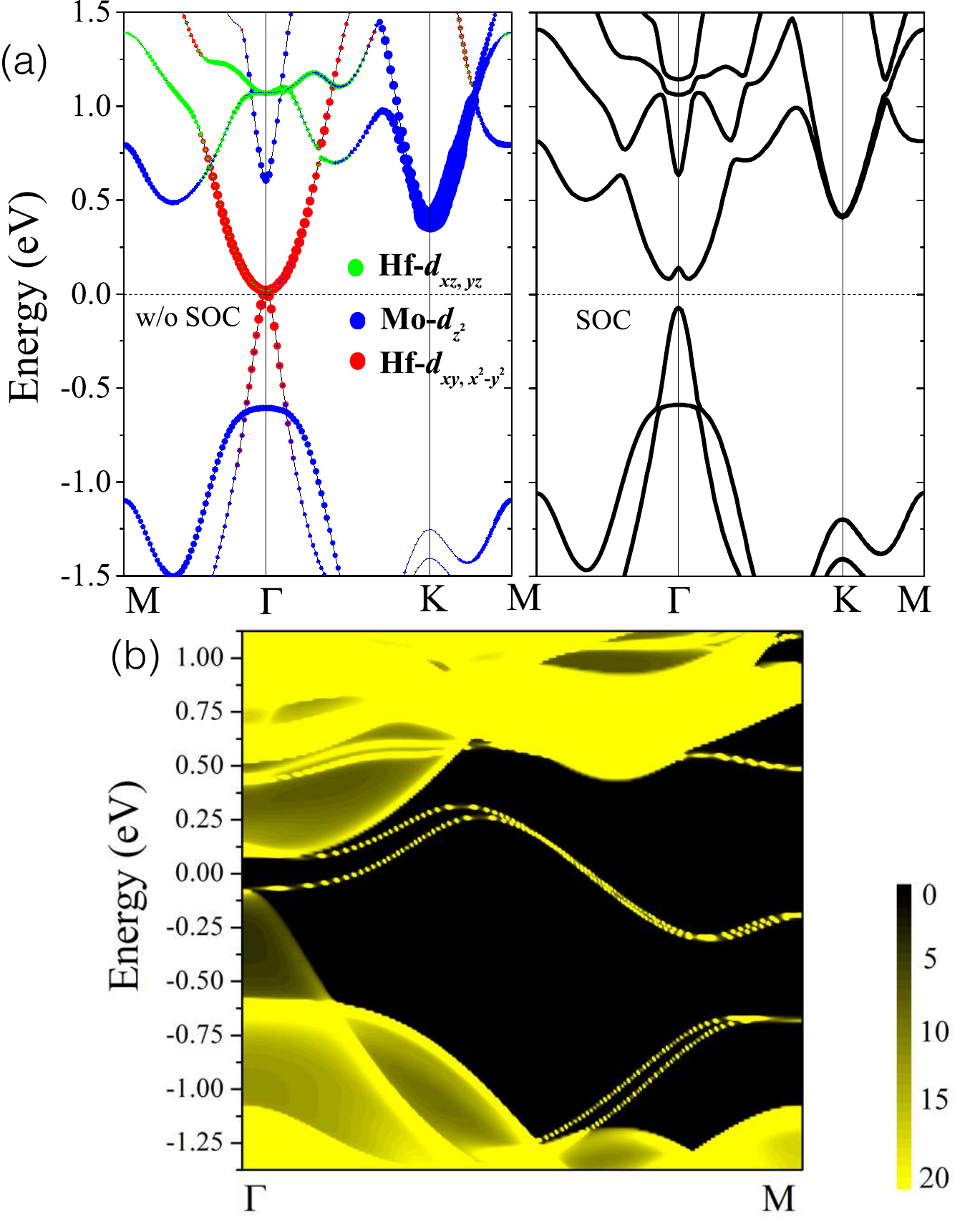}
  \caption{(a) Band structures of Mo$_2$HfC$_2$O$_2$
  without and with the SOC~\cite{C.Si2016}. (b) Edge states of nanoribbon of Mo$_2$HfC$_2$O$_2$ with zigzag edges~\cite{C.Si2016}.
 $\Gamma$ (0,0,0), M(1/2,0,0), K(1/3,1/3,0), X(1/2,1/2) are  
high-symmetry points of the Brillouin zone of a hexagonal structure. 
Fermi energy is located at zero energy.
}
  \label{fig:electronic1}
\end{figure}

 \begin{figure}[t]
\centering
\includegraphics[width=0.80\columnwidth]{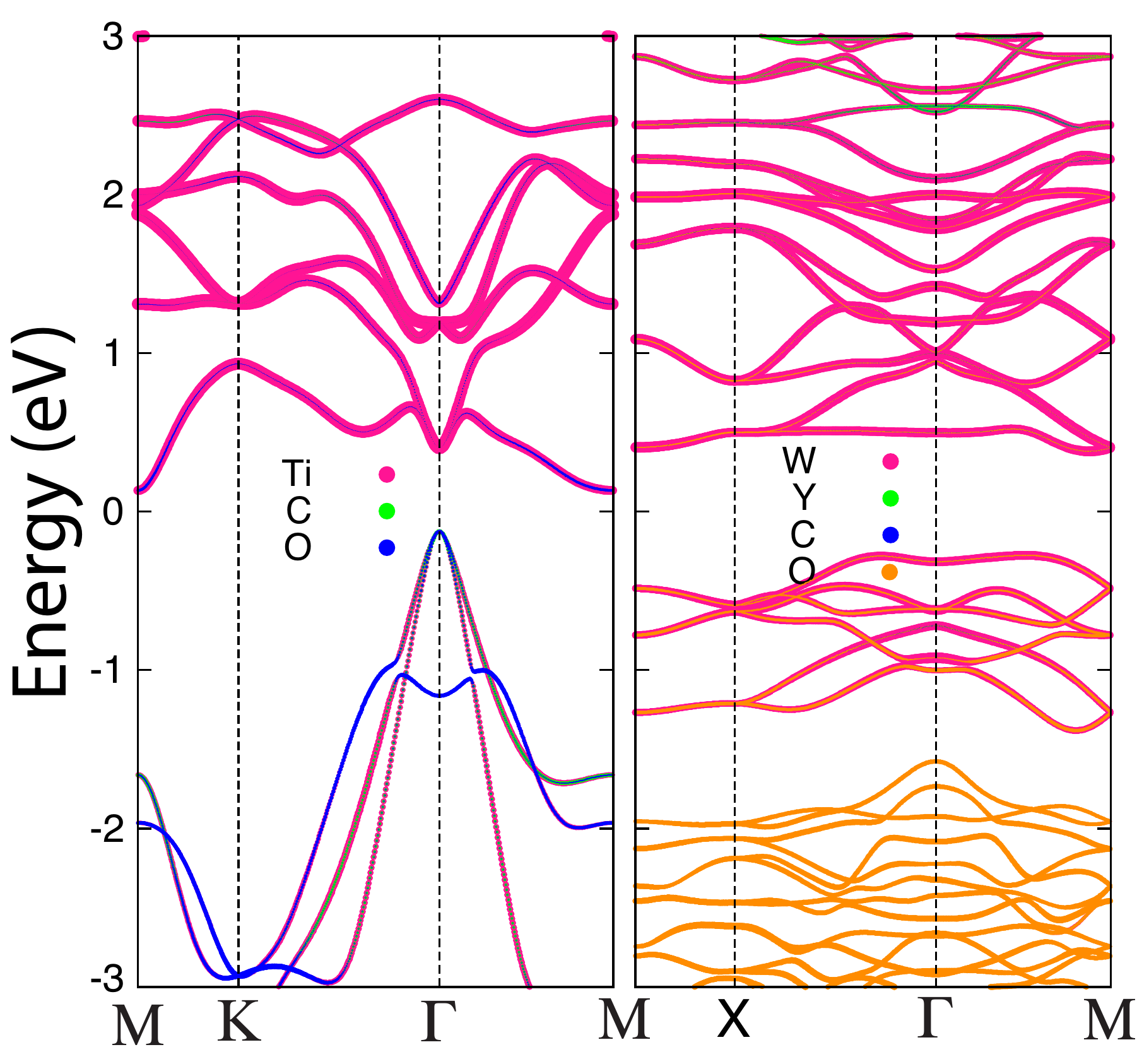}
  \caption{Left and right panels exhibit the projected band structures of Ti$_2$CO$_2$ with hexagonal lattice and (W$_{2/3}$Y$_{1/3}$)$_2$CO$_2$ with rectangular lattice. 
  $\Gamma$ (0,0,0), M(1/2,0,0), K(1/3,1/3,0), X(1/2,1/2) are  
high-symmetry points of the Brillouin zone. 
Fermi energy is located at zero energy.
}
  \label{fig:electronic2}
\end{figure}

Although the adsorption energies of F, O, and OH are large negative values and the mechanical properties improve 
after surface functionalization, 
the formation of functionalized MXenes depends on the 
competitive bulk phases. The calculation of the formation energies for M$_{n+1}$X$_n$O$_2$ 
($n=1$, 2, and 3; X = C or N)
shows that all formation energies are positive values and V$_2$CO$_2$ has the highest positive value (+0.285 eV/atom) among the
synthesized MXenes~\cite{M.Ashton2016_2}. 
This can be considered as a threshold for the formation of new MXenes. In other words, only 
the MXenes with formation energies less than that for V$_2$CO$_2$ have a chance for synthesis. 
Another important issue is the effect of the
degradation of MXenes over time and due to the temperature,  
which might result in the transformation of MXenes into bulk transition-metal carbides
or oxides~\cite{P.Srivastava2016,M.Naguib2014}. As an example, it has been experimentally observed 
that when 2D Ti$_3$C$_2$ is heated, it is transformed into TiO$_2$ particles enmeshed in thin sheets of graphitic carbon structures~\cite{M.Naguib2014}. 
Thermodynamic analyses show that the large concentration of HF might convert Ti$_3$C$_2$ to TiF$_3$ and TiF$_4$~\cite{P.Srivastava2016}.

\subsection{Electronic structures}

Electronically, MXenes can be either topologically nontrivial or trivial. 
Few funcionalized MXenes have been demonstrated to be topologically nontrivial semiconductors, e.g., 
M$_2$CO$_2$ (M = Mo, W), M$'_2$M$''$CO$_2$ (M$'$ = Mo, W; M$''$ = Ti, Zr, Hf), Ti$_3$N$_2$F$_2$, or semimetals, 
e.g., M$'_2$M$_2''$CO$_2$ (M$'$ = Mo, W; M$''$ = Ti, Zr, Hf)~\cite{H.Weng2015_1,M.Khazaei2016_2,L.Li2016,C.Si2016,C.Si2016_2,Y.Liang2017}. 
In these MXenes, 
the spin$-$orbit coupling (SOC) affects the electronic structures significantly.  
In more detail,
without including the SOC effect, the above MXenes are semimetals in which the 
highest valence band and lowest conduction band touch at the $\Gamma$ point at the Fermi 
energy. These bands are dominated with the $d$ orbitals of the transition metals.
Upon including the SOC effect, their degeneracy at the Fermi energy is lifted, and a 
gap opens at the $\Gamma$ point. For example, the band structures of nontrivial semiconducting Mo$_2$HfC$_2$O$_2$
MXene are shown in Fig.~\ref{fig:electronic1}.
They are called 2D topological semimetals and insulators, respectively, and 
their nanoribbon structures possess topologically protected conducting edge states that cross the Fermi energy. 
In other words, they display conducting edge states in which electrons with opposite 
spins propagate in opposite directions, and these edge states are robust against nonmagnetic impurities and 
disorder~\cite{M.Z.Hasan2010}. The nontrivial band topologies of these MXenes have also been confirmed 
by further band structure analyses and edge state calculations of their nanoribbon structures. The edge state of a zigzag nanoribbon 
structure of Mo$_2$HfC$_2$O$_2$ is shown in Fig.~\ref{fig:electronic1}, displaying the valence and conduction bands form a single Dirac cone point  at M point near the Fermi energy. 
It is noteworthy that some MXenes such as Sc$_2$C(OH)$_2$ turn into topological insulators with the application of an electric field and charge transfer~\cite{E.Balci2018}.

The majority of MXenes belong to the family of trivial metals, semimetals, or semiconductors.
In other words, the SOC does not change their electronic structures significantly.  
Almost all pristine MXenes are metallic. 
However, upon surface functionalization,
 some of them become semiconducting, e.g., Sc$_2$CT$_2$ (T= F, OH, O), M$_2$CO$_2$ (M = Ti, Zr, Hf)~\cite{M.Khazaei2013}, 
 and (M$'_{2/3}$M$''_{1/3}$)$_2$CO$_2$ (M$'$ = Mo, W; M$''$ = Sc, Y)~\cite{M.Khazaei2018_2}. 
 These MXenes become 
 semiconducting owing to the shift of the Fermi level and/or the change of the local crystal field 
 around the transition metals. For example, the pristine M$_2$X (X = C, N) systems are all metallic 
 with the Fermi energy located at the $d$ bands of the transition metals. 
 In most MXenes, the $p$ bands of C/N are below the $d$ bands of the transition metals and these 
 bands are separated by a small band gap~\cite{M.Khazaei2013}. 
F, OH, or O functionalization generates new bands 
below the Fermi energy, hybridized with the M $d$ orbitals. 
 Sc$_2$CT$_2$ (T= F, OH, O) and M$_2$CO$_2$ (M = Ti, Zr, Hf) become semiconducting because 
 the Fermi energy is located at the center of the gap between the M $d$ bands and the X $p$ bands after functionalization~\cite{M.Khazaei2013}. 
 The band structure of Ti$_2$CO$_2$ is shown in Fig.~\ref{fig:electronic2}.
 
\begin{figure}[t]
\centering
\includegraphics[width=1.0\columnwidth]{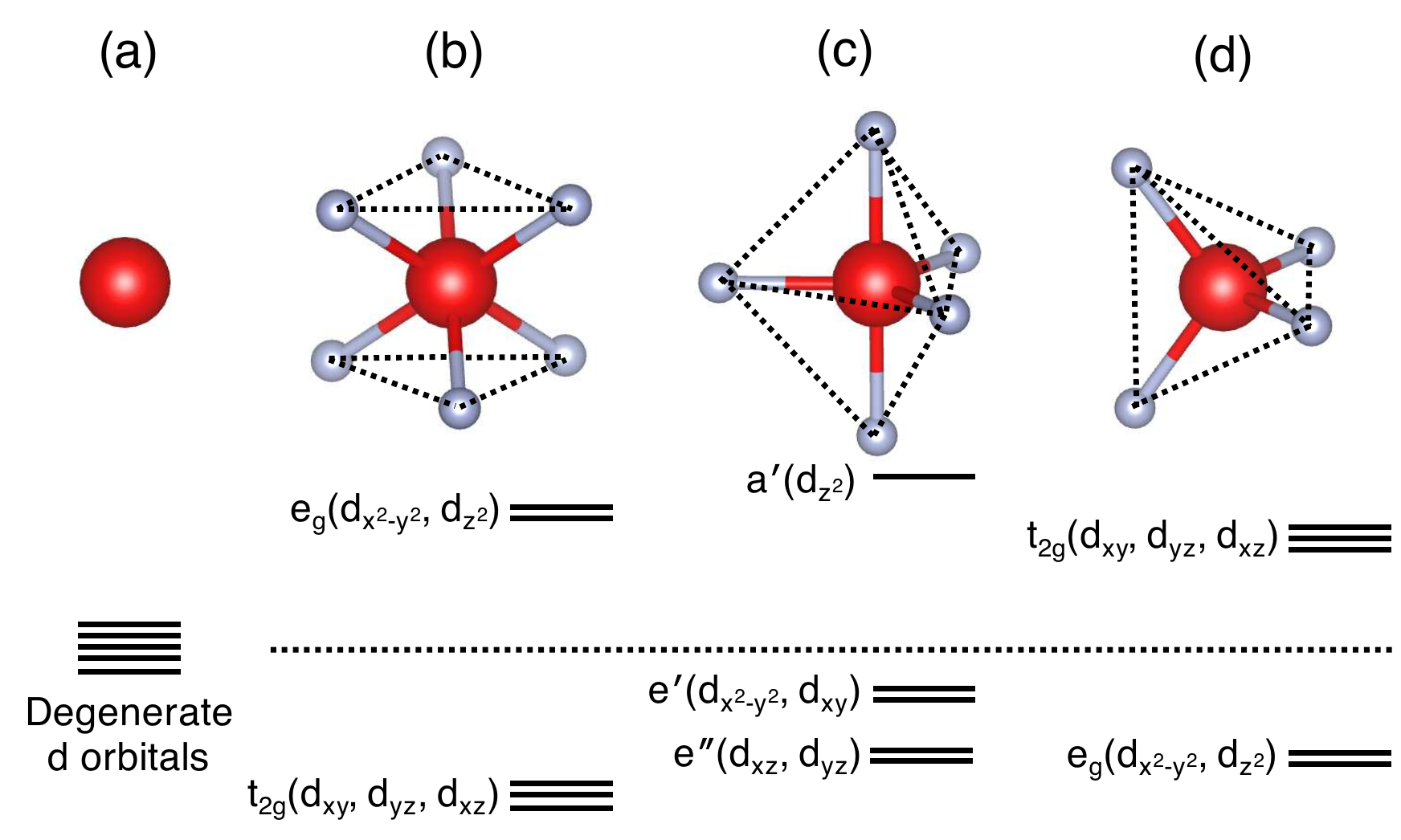}
  \caption{Schematic energy levels for $d$ orbitals of a transition metal (a) 
  and those under octahedral (b), trigonal bipyramidal (c), 
  and tetrahedral (d) coordination environments, which are usually found in 
  two-dimensional structures. A red (grey) sphere denotes a transition metal (ligand). 
}
  \label{fig:splitting}
\end{figure}

In the case of (M$'_{2/3}$M$''_{1/3}$)$_2$C (M$'$ = Mo, W; M$''$ = Sc, Y), the Fermi energy moves to lower 
energies upon functionalization with oxygen. The origin of the band gaps in (M$'_{2/3}$M$''_{1/3}$)$_2$CO$_2$ is 
the splitting of $d$-orbital bands~\cite{M.Khazaei2018_2} due to the crystal field around the transition metals 
surrounded by C and O atoms~\cite{T.Hu2017}. Schematics of $d$-band splittings by various types of crystal fields are shown in Fig.~\ref{fig:splitting}.
In more detail, 
the states near the Fermi energy are t$_{2g}$ bands hybridized with C and O $p$ orbitals, which split widely, resulting in finite band gaps~\cite{M.Khazaei2018_2}. 
As an example, the band structure of (W$_{2/3}$Y$_{1/3}$)$_2$CO$_2$ is
shown in Fig.~\ref{fig:electronic2}.

It is noteworthy that a \texttt{aNANt} database containing the structural and electronic information of approximately  15,000 MXenes has recently been released~\cite{anant}. The chemical formula considered in this database is MM$^\prime$XTT$^\prime$, where M/M$^\prime$ is an early transition metal, X is C or N, and T/T$^\prime$ stands for 14 different termination groups such as F, OH, CN, and SCN~\cite{A.C.Rajan2018}. The electronic properties have been predicted with a combination of DFT and machine learning. It has been found that as the electronegativity difference between the functional groups and the transition metal becomes larger, there is a higher chance for semiconducting MXenes~\cite{A.C.Rajan2018}. It should also be noted that semiconducting 
MXenes can be engineered using strain~\cite{X.F.Yu2015_2}, electric field~\cite{Y.Lee2014}, or by being placed 
on the other 2D systems~\cite{Y.Lee2015}.

In brief, the chance for obtaining semiconducting MXenes is higher for thin MXenes 
($n = 2$; M$_2$XT$_2$ or (M$'_{2/3}$M$''_{1/3}$)$_2$XT$_2$) than for thicker ones ($n =  3$ and 4). 
All M$_3$X$_2$T$_2$ and M$_4$X$_3$T$_2$ MXenes are metallic. In the family of ordered double transition metals, some of M$'_2$M$''$C$_2$ could become semiconducting because of spin-orbit interaction or 
magnetic ordering as will be discussed in the next section. 
Most of M$'_2$M$_2''$C$_3$ are metallic except for few of them that are semi-metal because of 
spin-orbit interaction. Generally, metallic or semiconducting behavior of multilayer MXenes is mostly inherited from 
their monolayers.

\subsection{Magnetic states}
 
 Almost all MAX phases that have been reported to be magnetic are made of Cr and/or Mn, e.g., 
 Cr$_2$AlC, Cr$_2$GeC, Cr$_2$GaC, Cr$_2$AlN, C$_2$GaN, Mn$_2$AlC, Mn$_2$GaC, Cr$_2$TiAlC$_2$, 
 and (Cr$_{2/3}$Sc$_{1/3}$)$_2$AlC with various magnetic orders~\cite{B.Anasori2015_2,J. Lu2017,A.S.Ingason2016}. 
Theoretically many pristine and/or functionalized MXenes with F, O, or OH have been predicted to be magnetic, e.g., M$_2$X (M = Ti, V, Cr, Mn; X = C, N), 
M$_2$MnC$_2$ (M = Ti, Hf), M$_2$TiX$_2$ (M = V, Cr, Mn; X =C , N), Hf$_2$VC$_2$, and Mo$_3$N$_2$F$_2$~\cite{G.Gao2016_1,C.Si2015,J.He2016,H.Kumar2017,L.Dong2017,W.Chen2017,S.S.Li2017,N.C.Frey2018,W.Sun2018,J.He2018,J.J.Zhang2018}.

The magnetic moments at the transition metals in the magnetic MXenes can mostly be determined 
from the nominal oxidation states of the transition metals,
C$^{4-}$, N$^{3-}$, F$^{-}$, OH$^{-}$, and O$^{2-}$, 
and also by examining the coordination number of the transition metals and the number of $d$ electrons~\cite{H.Kumar2017,L.Dong2017}.
Similar to dichalcogenides, the nonbonding $d$ orbitals in MXenes are formed near the Fermi level 
located between the bonding and antibonding states. 
Therefore, only the electrons that occupy the nonbonding $d$ orbitals can  
mainly contribute to the magnetism.

As explained above, in functionalized MXenes, each transition metal is surrounded by C/N atoms and termination chemical groups, which form an octahedral cage around the transition metal, see Fig.~\ref{fig:splitting}. 
The resulting nearly octahedral crystal field splits the $d$ orbital of the transition metal 
into the $t_{2g}$ ($d_{xy}$, $d_{yz}$, and $d_{xz}$) and $e_g$ ($d_{x^2-y^2}$ and $d_{z^2}$) orbitals. 
Because of the orbital shapes, the $e_g$-orbital manifold is energetically higher than the $t_{2g}$-orbital 
manifold. Therefore, electrons occupy first the $t_{2g}$ orbitals before entering the $e_g$ orbitals. 
 Various magnetic behaviors of MXenes are expected because of
the different numbers of available electrons of the transition metals and the different electron configurations of 
the $d$ orbitals~\cite{H.Kumar2017}. For example, a Cr atom has six valence electrons.  The nominal oxidation 
state of Cr is $+3$ in Cr$_2$CF$_2$ because, in this structure, each Cr atom donates two electrons to the neighboring 
C atom and one electron to the neighboring F atoms. Therefore, three electrons remain on each Cr atom that, according to Hund's rule, fill the t$_{2g}$ orbitals with the maximum spin, generating a magnetic moment of $3\mu_B$. 
Depending on how strongly the majority and minority spin bands split, the magnetic MXenes become metallic, semiconducting, or a half-metal~\cite{G.Gao2016_1,C.Si2015,J.He2016,H.Kumar2017,L.Dong2017,W.Chen2017,N.C.Frey2018}. 
For example, Cr$_2$CF$_2$ is a semiconducting MXene~\cite{H.Kumar2017}.

\begin{figure}[t]
\centering
\includegraphics[width=1.0\columnwidth]{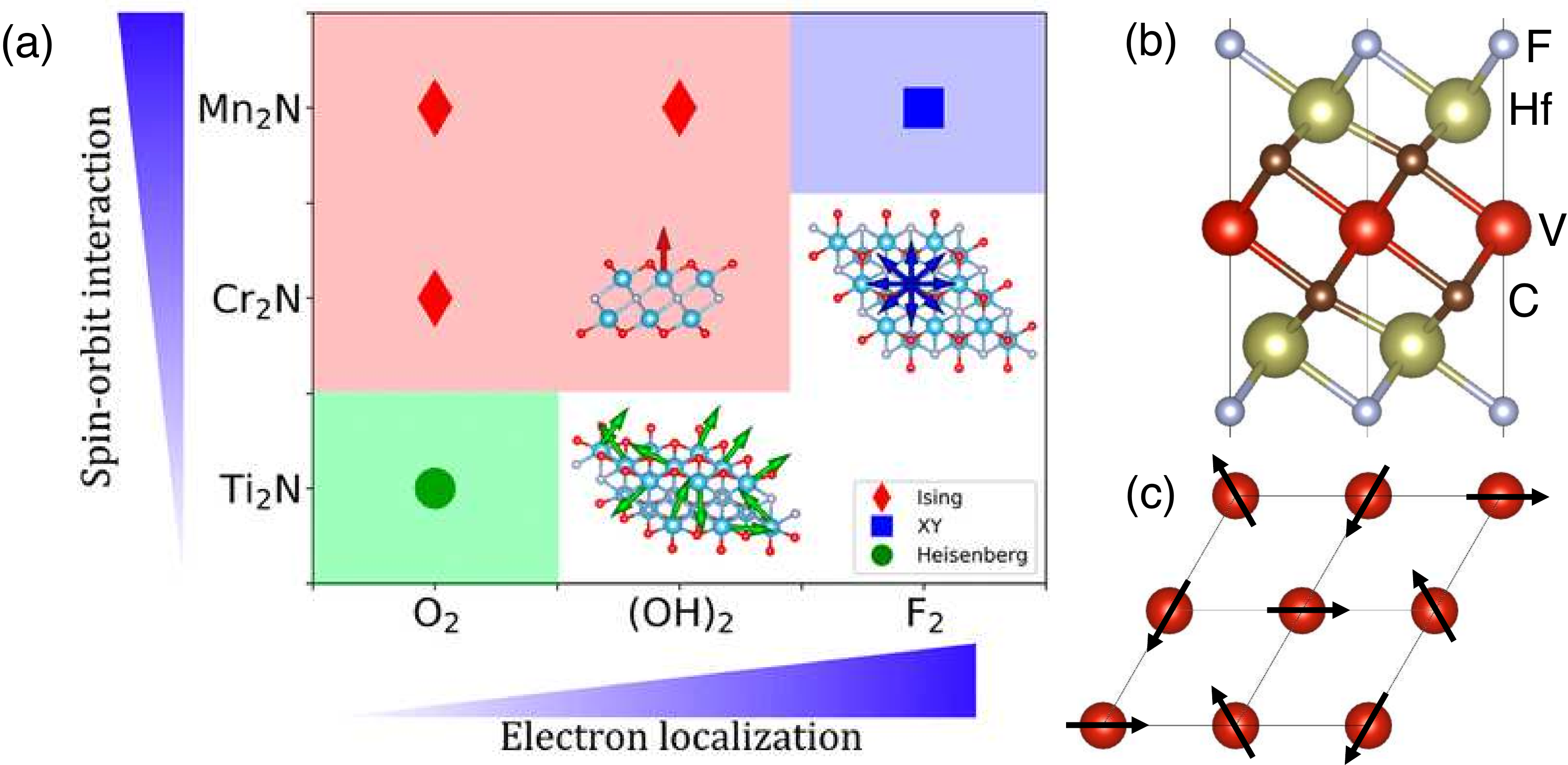}
  \caption{(a) Various spin states of M$_2$N (M = Ti, Cr, Mn) according to spin-orbit coupling and electron localization effect~\cite{N.C.Frey2018}. 
  (b) Side view of 2D Hf$_2$VC$_2$F$_2$. (c) Top view of the $ab$ lattice structure formed by V atoms only 
  in 2D Hf$_2$VC$_2$F$_2$. Arrows indicate the spin patterns at V atoms, displaying 
120$\degree$ noncollineanr anti-ferromagnetic order~\cite{J.J.Zhang2018}.  
}
  \label{fig:magnetic}
\end{figure}

Recent systematic calculations on M$_2$N (M = Ti, Cr, Mn) found that various magnetic orders (ferro, anti-ferro, or ferri)  
with different magnetic interactions (Ising, XY, or Heisenberg type) can be controlled through the strength of spin-orbit 
coupling, 
using different transition metals M, and the degree of localization of electrons, changed via the surface termination 
with electronegative elements (e.g. F and O) or chemical groups (e.g. OH)~\cite{N.C.Frey2018}, as schematically 
shown in~Fig~\ref{fig:magnetic}(a). 
Among the magnetic MAX phases, only Cr$_2$TiAlC$_2$ has 
been exfoliated into Cr$_2$TiC$_2$~\cite{B.Anasori2015_2}. Interestingly, all members of Cr$_2$TiC$_2$T$_2$ (T= H, F, O, OH) 
are semiconductors independently of the termination group~\cite{W.Sun2018}. It is noteworthy that the majority of magnetic MXenes have 
anti-ferromagnetic ground states except for some of the Mn-based ones, which are ferromagnetic~\cite{W.Sun2018}. 
Furthermore, 
the magnetic and electronic states of M$_2$X MXenes can be engineered by strain~\cite{G.Gao2016_1}.
More recently, as shown in Figs.~\ref{fig:magnetic}(b) and \ref{fig:magnetic}(c), due to the spin-orbit coupling,  
Hf$_2$VC$_2$F$_2$ was predicted to exhibit in-plane noncollinear 120$\degree$ magnetic order~\cite{J.J.Zhang2018}.

\subsection{Photonic properties}

Various optical properties of materials such as the transmittance, absorption, reflectivity, refractive index, 
and energy loss can be predicted theoretically through an investigation of the real and imaginary parts of the 
complex dielectric constant, 
$\epsilon(\omega)=\epsilon_1(\omega)+i~\epsilon_2(\omega)$, as a function of the photon wavelength or 
frequency $\omega$. 
The real part $\epsilon_1(\omega)$ [imaginary part $\epsilon_2(\omega)$] of the optical dielectric function is 
obtained using the Kramers$-$Kronig transformation (via the sum of the interband transitions from 
the occupied to unoccupied states allowed by the electric-dipole interaction)~\cite{P.Ravindran1999}.
For instance, the photon absorption coefficient is evaluated as 
$I(\omega)=\sqrt{2}\omega[\sqrt{\epsilon_1(\omega)^2 +\epsilon_2(\omega)^2}-\epsilon_1(\omega)]^{1/2}$, 
indicating that larger $\epsilon_2(\omega)$ results in better photon absorption.

\begin{figure}
\centering
\includegraphics[width=0.85\columnwidth]{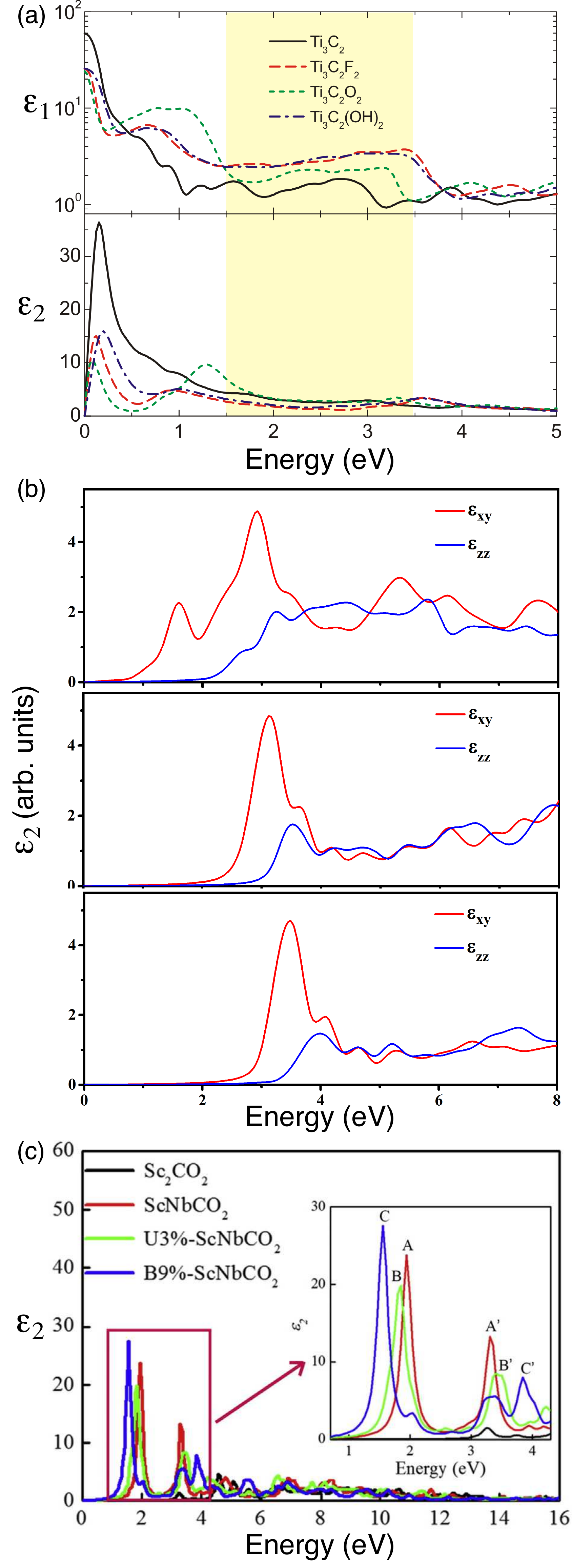}
\caption{(a) Real ($\epsilon_1$) and imaginary ($\epsilon_2$) parts of the dielectric constant versus frequency $\omega$ 
for pristine and F-, OH-, and O-functionalized Ti$_3$C$_2$ MXenes~\cite{G.R.Berdiyorov2016}. 
(b) In- and out-of-plane $\epsilon_2(\omega)$ (indicated by $\epsilon_{xy}$ and $\epsilon_{zz}$, respectively) 
for O-functionalized Ti, Zr, and Hf MXenes~\cite{H.Zhang2016}. 
(c) $\epsilon_2(\omega)$ for Sc$_2$CO$_2$ and Nb-doped Sc$_2$CO$_2$ without and with 3$\%$ uniaxial 
and 9$\%$ biaxial strain~\cite{J.Guo2017}.}
\label{fig:optical}
\end{figure}

As the electronic characteristics of MXenes change upon surface termination, their optical properties are accordingly affected. For example, as shown in 
Fig.~\ref{fig:optical}(a), upon functionalization of Ti$_3$C$_2$ with F, OH, and O, the real part $\epsilon_1(\omega$ = 0) of the static dielectric constant 
slightly decreases, but it does not change drastically. At $\omega\neq 0$, however, $\epsilon_1(\omega)$ increases with functionalization~\cite{G.R.Berdiyorov2016}. 
As seen in Fig.~\ref{fig:optical}(a), O-terminated Ti$_3$C$_2$ exhibits prominent peaks at frequencies below those of visible light for $\epsilon_1(\omega)$ as compared to the F- and 
OH-functionalized MXenes. Further, because of the similar electronic structures of F- and OH-terminated Ti$_3$C$_2$, 
$\epsilon_2(\omega)$ exhibits similar behavior at all energies~\cite{Y.Bai2016}. It is observed that in the visible light spectrum, the light absorption ability of O-terminated 
Ti$_3$C$_2$ is relatively better, i.e., larger $\epsilon_2(\omega)$, than that of pristine, F-terminated, or OH-terminated MXenes. This is attributed to the formation of O states close the Fermi energy.

Further calculations reveal that the dielectric function tensor of MXenes is anisotropic along the in-plane 
and out-of-plane direction, analogs to their crystal structure. As shown in Fig.~\ref{fig:optical}(b), the calculated $\epsilon_2(\omega)$ for O-functionalized M$_2$C (M=Ti, Zr, Hf) with semiconducting properties~\cite{M.Khazaei2013} displays peaks in the visible light region. Among these MXenes, Ti$_2$CO$_2$ with two peaks at energies around 1.5 and 3.0 eV possesses the highest absorption efficiency~\cite{H.Zhang2016}, and thus can be considered as a promising candidate for applications in optical devices. 
However, Ti$_2$CO$_2$ does not absorb light below 1 and 2.3 eV in the in- and out-of-plane directions, respectively, because of the different reflectivities of 12$\%$ and 6$\%$. 
Interestingly, the reflectivity is affected by the layer thickness and is higher for Ti$_3$C$_2$T$_2$ (T = F, OH, O) as compared to Ti$_2$CT$_2$~\cite{Y.Bai2016}.

Similar to O-functionalized M$_2$C (M=Ti, Zr, Hf), Sc$_2$CO$_2$ is also a semiconductor. 
As shown in Fig.~\ref{fig:optical}(c), $\epsilon_2(\omega)$ of
 the Sc$_2$CO$_2$ has a small peak at 3.3 eV (the same value as its band gap), which results from the electron transition from the C $p$ orbitals  to the Sc $d$ orbitals.  By replacing one Sc atom with a Nb atom (ScNbCO$_2$), a new peak appears at 1.9 eV, which is due to the reduction of the band gap as large as 1.89 eV~\cite{J.Guo2017}. Furthermore, the positions of the absorption peaks can be tuned by applying stress 
[see Fig.~\ref{fig:optical}(c)]. It is also shown that Sc$_2$CO$_2$ exhibits in- and out-of-plane polarization anisotropy, 
implying the application of the O-functionalized Sc-based MXenes for newly developed polarization-driven photovoltaics~\cite{A.Chandrasekaran2017}.

Recently, many photonic-based applications have been developed for MXenes~\cite{K.Hantanasirisakul2018}. 
For example, some MXenes such as V$_2$CT$_x$ (T = F, OH, O) with having absorption in the range of 
500 to 2700 nm and high conductivity have been utilized as conductive transport electrodes~\cite{G.Ying2018}. 
Ti$_3$C$_2$T$_x$ with almost full (84$\%$) light-to-water evaporation efficiency has the potential for application 
in photothermal evaporation systems~\cite{R.Li2017}. 
Some MXenes show nonlinear light absorption applicable for optical switching devices~\cite{K.Hantanasirisakul2018}.

 \begin{figure}[t]
\centering
\includegraphics[width=1.0\columnwidth]{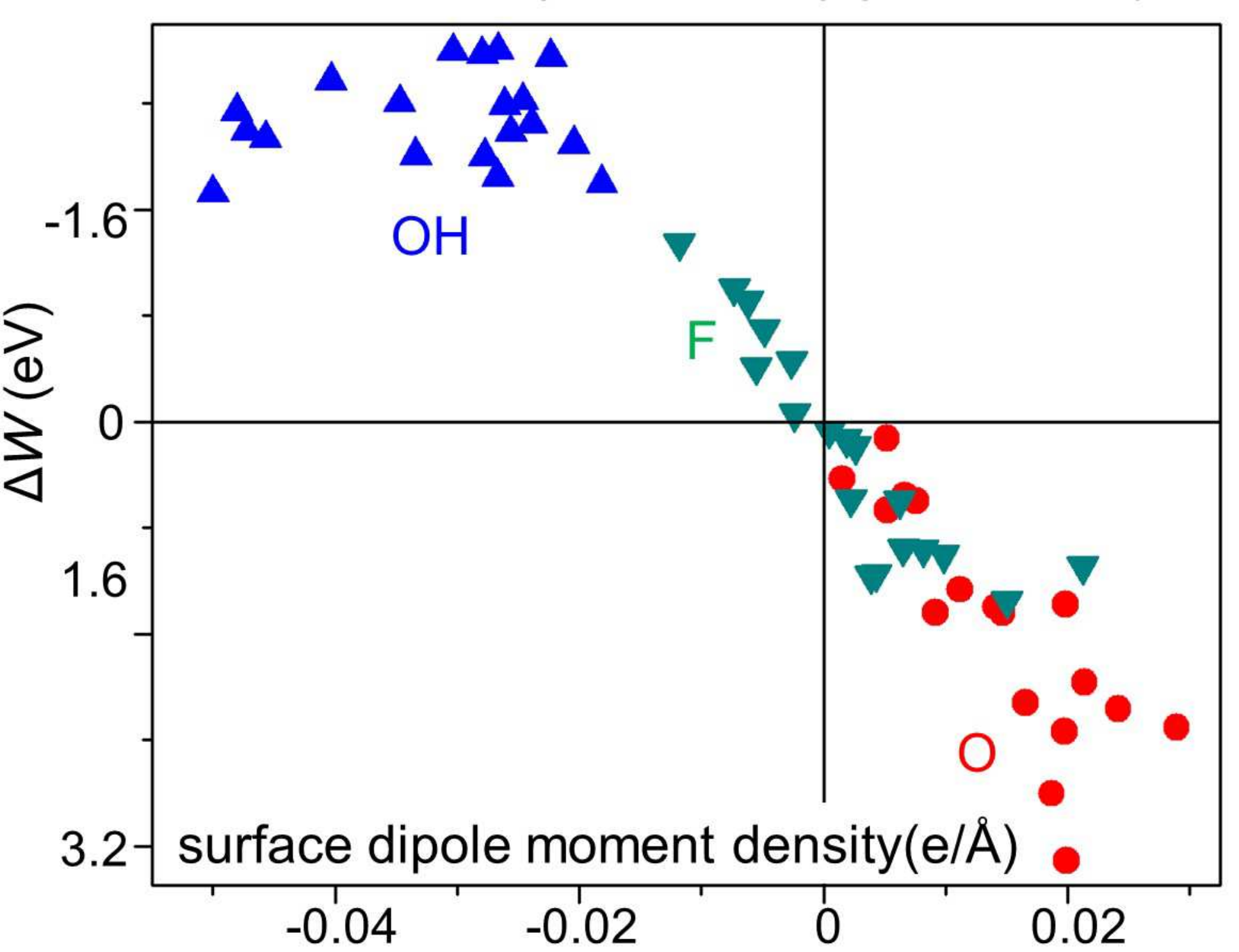}
  \caption{The effect of surface functionalization on the change of work function and surface dipole moment~\cite{Y.Liu2016}.    
}
  \label{fig:workfunction}
\end{figure}

\subsection{Surface properties}
 
The work function, defined as the difference between the Fermi level and
the vacuum potential, is one of the useful surface characteristics of materials 
that are also used for the purpose of device applications. 
The work 
functions of pristine MXenes and MXenes functionalized with F, OH, and O are estimated to be in 
the ranges of 3.3$-$4.8, 3.1$-$5.8, 1.6$-$2.8, and 3.3$-$6.7 eV, respectively~\cite{M.Khazaei2015}. 
OH-terminated MXenes can be considered as ultralow work function materials.
The change in the work function of MXenes after functionalization with respect to pristine MXenes, $\Delta$W, can be 
explained in terms of the change in the surface dipole moments ($\Delta$P)~\cite{M.Khazaei2015,Y.Liu2016}. 
It is found that $\Delta$W and $\Delta$P have a linear correlation~\cite{M.Khazaei2015,Y.Liu2016} (see Fig.~\ref{fig:workfunction}). 
 The change in the surface dipole moments after surface functionalization results from    
 i) the charge transfer and redistribution between the surface and the chemical groups, 
 ii) the structural relaxation caused by the chemical groups, and iii) the polarity of the 
 group~\cite{M.Khazaei2015}. Considering their work function properties, MXenes 
can be used to construct Schottky-barrier-free contacts with other 2D semiconductors,
in which all OH-terminated (some O-terminated) MXenes can act as electron (hole) injectors~\cite{Y.Liu2016}.
It is noted that the above trends in the work functions of functionalzied MXenes are independent of 
their thickness ($n=1$--3)~\cite{M.Khazaei2015}.
Interestingly, the recent studies on Ti$_3$C$_2$O$_x$F$_y$(OH)$_z$ with mixtures of surface terminations have 
shown that the ultralow work function of fully OH terminated MXenes can be attained by 60$\%$ OH termination 
on the surface~\cite{N.M.Caffrey2018}.

 \begin{figure}[t]
\centering
\includegraphics[width=0.7\columnwidth]{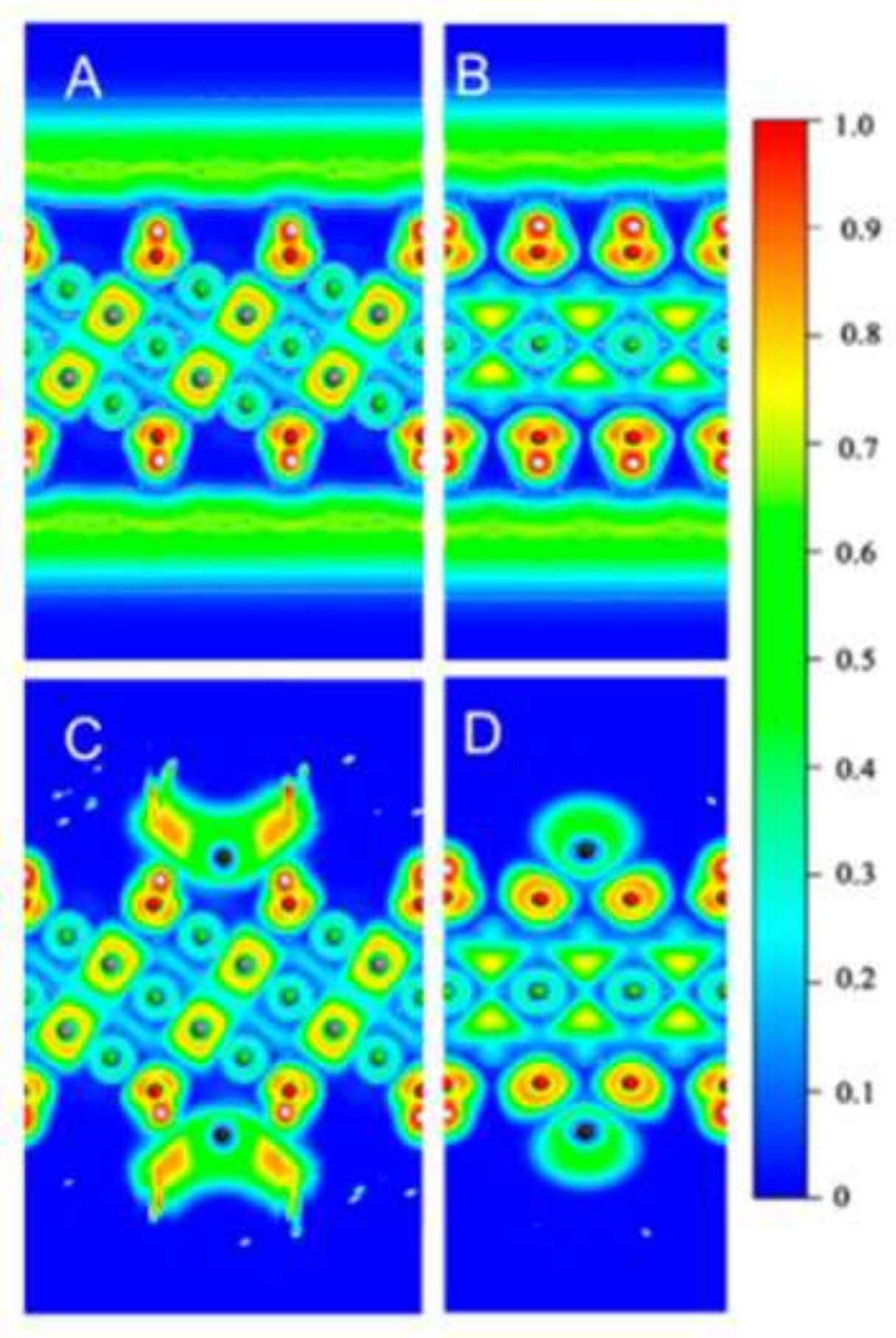}
  \caption{Electron localization function (ELF) counterplots of Ti$_3$C$_2$O$_2$H$_{2(1-m)}$Pb$_m$ 
  for 1/9 monolayer Pb impurity~\cite{Q.Peng2014}. 
 (A) and (B) display ELF for pristine Ti$_3$C$_2$(OH)$_2$ on $(110)$ and $(-110)$ planes, respectively. 
 (C) and (D) show ELF for Ti$_3$C$_2$O$_2$H$_{2(1-m)}$Pb$_m$ on $(110)$ and $(-110)$ planes, respectively.    
}
  \label{fig:ELF}
\end{figure}

 Like many other 2D systems, e.g., graphene, BN, and MoS$_2$, nearly free electron (NFE) states appear 
 in MXenes in the spatial region above the surfaces toward the vacuum~\cite{M.Khazaei2016_1}. 
 These states have parabolic bands with respect to the wave vector and generally appear at high energies above the Fermi energy and generally are not accessible except 
 by applying electric field. 
 Energetically, the NFE states in O- and F-terminated MXenes also appear at high energies above the Fermi energy. 
However, in the case of the OH termination, the NFE states appear near the Fermi energy, and for some of the OH-terminated MXenes, they are 
partially occupied by electrons. 
Because of the positively charged nature of the H atoms in OH-terminated MXenes, 
their surface potentials are 
shallow with an extended tail at energies near the Fermi energy, but they are very deep in O- or F-terminated MXenes~\cite{M.Khazaei2016_1}. 
Hence, the NFE states in OH-terminated (F- or O-terminated) MXenes can be found at energies near (above) the Fermi energy.  

The NFE states are formed spatially above the surface and do not have any weight when they are projected 
onto the atomic orbitals of the compositional elements. 
However, the existence of partially occupied NFE states can be observed 
by visualizing the electron localization function (ELF).
Figure~\ref{fig:ELF} shows that a uniform electron gas exists on top of the hydrogen atoms (ELF = 0.4) above Ti$_3$C$_2$(OH)$_2$. 
 This implies that the OH-terminated MXenes
with partially occupied NFE states can be highly sensitive to
adsorbents. Therefore, they can be used for selective gas
sensors~\cite{M.Khazaei2016_1} and heavy-element purification applications~\cite{Q.Peng2014}. 
For example, as shown in Fig.~\ref{fig:ELF}, upon Pb atom adsorption, 
the NFE states of Ti$_3$C$_2$(OH)$_2$ disappear~\cite{Q.Peng2014}.
It is interesting to note that the location of NFE states relative to the Fermi energy can be tuned 
by either surface charges~\cite{N.Alidoust2014} 
or applied electric filed~\cite{M.Otani2010}.
NFE states can contribute to electron transport~\cite{M.Khazaei2016_1} and surface catalytic properties. 
NFE states can be observed by X-ray photoelectron spectroscopy measurements~\cite{N.Alidoust2014}. 
The location of NFE states in energy is almost independent of the thickness of MXenes ($n=1$--3). 
The NFE states on OH-terminated MXenes disappear when the OH-terminated MXenes are sandwich between other 2D systems such as graphene and multilayer MXenes because the surface potential changes.

\begin{figure}[t]
\centering
\includegraphics[width=0.93\columnwidth]{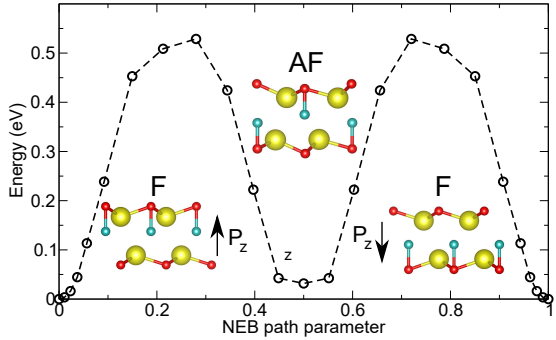}
  \caption{The calculated energy barrier using nudged elastic band (NEB) method for switching polarization of ferroelectric Sc$_2$CO$_2$
from +P$_z$ to $-$P$_z$~\cite{A.Chandrasekaran2017}. The $z$-axis is normal to the surface. 
The ferroelectric (F) states of Sc$_2$CO$_2$ are switched through an intermediate antiferroelectric (AF) 
state, in which the polarization is almost zero.  The arrows indicate the direction of the polarizations.
}
  \label{fig:ferroelectric}
\end{figure}

\section{Applications of MXenes}

\subsection{Ferroelectricity}
Noncentrosymmetric 2D semiconductors can exhibit ferroelectricity and/or piezoelectricity owing to their polar structures.
A ferroelectric material is a polar system whose polarity can be reversed by applying an electric field.   
 Among MXene family members, Sc$_2$CO$_2$ has uniquely in- and out-of-plane polarizations. This is due to the asymmetric occupation of 
 O atoms on the surfaces of Sc$_2$C; on one side of the surfaces, 
 they are adsorbed at the hollow site at which no C atoms available under them and on the other side of the surfaces, the O atoms are adsorbed at hollow sites at which C atoms are available under them. 
Sc$_2$CO$_2$ possesses an out-of-plane polarization of 1.60 $\mu$C/cm$^2$ with respect to its nonpolar phase~\cite{A.Chandrasekaran2017}. The height of the energy barrier for switching from the ferroelectric phase to the anti-ferroelectric phase
is 0.53 eV per formula unit, as shown in Fig.~\ref{fig:ferroelectric}, which is sufficiently high for room-temperature device applications~\cite{A.Chandrasekaran2017}. Interestingly, 
although Sc$_2$CO$_2$ is a semiconductor with a band gap of 1.91 eV, bilayer is of this material is a semimetal, where 2D electron and hole gases exist on the opposite sides of the surfaces~\cite{A.Chandrasekaran2017}.  

 More recently, it has also been reported theoretically that multiferroelectricity can be achieved in 
ordered-double transition metal MXenes, e.g., in Hf$_2$VC$_2$F$_2$, where
 the helical spin states [see Figs.~\ref{fig:magnetic}(b) and \ref{fig:magnetic}(c)] generate an electrical 
 polarization on the surface~\cite{J.J.Zhang2018}.

 \begin{table}
\center
\caption{
Calculated relaxed-ion piezoelectric stress ($e_{11}$ in $10^{-10}$ Cm$^{-1}$) and 
strain ($d_{11}$ in pVm$^{-1}$)
coefficients for Sc$_2$CO$_2$ and (M$'_{2/3}$M$''_{1/3}$)$_2$CO$_2$ (M$'$ = Mo, W; M$''$ = Sc, Y) iMXenes~\cite{M.Khazaei2018_2}.}
\label{tab:piezo}
\begin{tabular}{lcccc}
\hline
\hline
 MXene            &                              $e_{11}$   &    $d_{11}$  &     \\
\hline
Sc$_2$CO$_2$                                      &  3.333  &  4.137 & \\
(Mo$_{2/3}$Sc$_{1/3}$)$_2$CO$_2$   & 44.55 &  29.24  &   \\         
(Mo$_{2/3}$Y$_{1/3}$)$_2$CO$_2$    & 40.33 &   35.91  &   \\
(W$_{2/3}$Sc$_{1/3}$)$_2$CO$_2$    & 38.82 &  21.67   &    \\
 (W$_{2/3}$Y$_{1/3}$)$_{2}$CO$_2$  & 35.53 &   24.98  &   \\                                                                                                                      
\hline
\hline
\end{tabular}
\end{table}

\subsection{Piezoelectricity}

Piezoelectrics are a family of semiconducting materials with an in-plane noncentrosymmetry for which 
an electrical field can be generated by a mechanical strain because of the shifts 
of the centers of positive and negative charges. 
The stress and strain piezoelectric properties of materials under uniaxial strain 
can be measured by $e_{11}$ and $d_{11}$, respectively. 
In the MXene family, Sc$_2$CO$_2$ and (M$'_{2/3}$M$''_{1/3}$)$_2$CO$_2$ (M$'$ = Mo, W; M$''$ = Sc, Y)
are semiconductors without in-plane centrosymmetry~\cite{M.Khazaei2018_2}. 
The piezoelectric coefficients of these materials are summarized in Table~\ref{tab:piezo}~\cite{M.Khazaei2018_2}.
It is found that $d_{11}$ for Sc$_2$CO$_2$ is in the same range of those for well-known transition metal dichalcogenides 
(MoS$_2$ and MoSe$_2$, 3.65 and 4.55 pVm$^{-1}$, respectively)~\cite{M.N.Blonsky2015}. 
However, $d_{11}$ for (M$'_{2/3}$M$''_{1/3}$)$_2$CO$_2$ (M$'$ = Mo, W; M$''$ = Sc, Y) is notably larger than those of many piezoelectric 2D materials such as BN, GaAs, GaSe, CaS, and AlSb 
(in the range of 0.5$-$3.0 pmV$^{-1}$~\cite{W.Li2015,C.Sevik2016}) and bulk materials including $\alpha$-quartz, wurtzite GaN,
and wurtzite AlN (2.3, 3.1, and 5.1 pmV$^{-1}$, respectively~\cite{R.Bechmann1958, C.M.Lueng2000}), which are generally used in industry.
This emphasizes that some semiconducting MXenes may find applications as piezoelectrics.

\begin{figure}
\centering
\includegraphics[width=0.6\columnwidth]{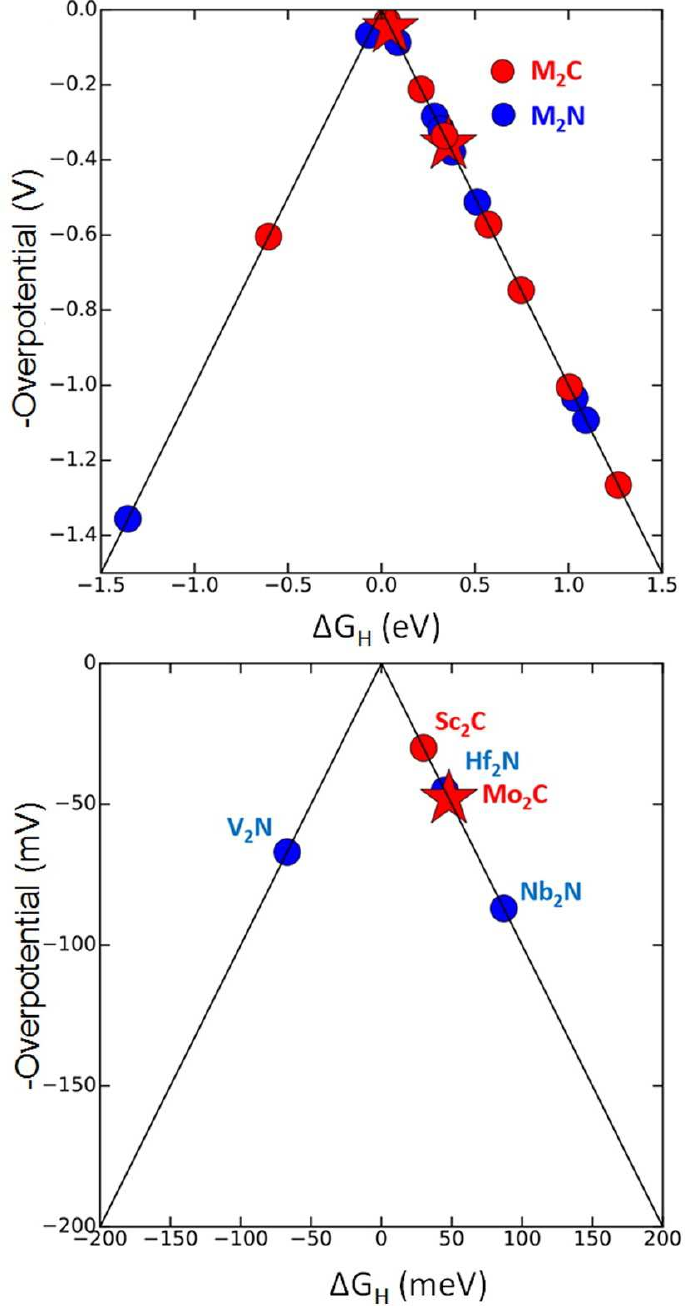}
\caption{Top panel: Volcano plot for the hydrogen evolution reaction (HER). Red and blue circles show carbide and nitride MXenes, respectively, whereas the experimentally known MXenes are indicated by stars~\cite{Z.Seh2016}. Bottom panel: Magnified version of the volcano plot.}
\label{fig:cata}
\end{figure}

\subsection{Thermoelectricity}

Thermoelectrics are a family of materials that can generate electricity by applying temperature gradient and vice-versa.
The thermoelectric efficiency  
is quantified through the dimensionless figure of merit ZT, expressed as $S^2\sigma T/K$, where $S$, $\sigma$, 
$T$, and $K\,(=k_{\rm l}+k_{\rm e})$ are the Seebeck coefficient, electrical conductivity, temperature, and 
thermal conductivity with both lattice ($k_{\rm l}$) and electronic ($k_{\rm e}$) contributions, respectively. 
The thermoelectric properties of various pristine and functionalized MXenes have already been examined theoretically, using a combination of Boltzmann theory and first-principles calculations~\cite{M.Khazaei2014_3}. Although metallic MXenes possess very good electrical conductivities, their Seebeck coefficients are almost negligible, resulting in a 
poor power factor ($S^2\sigma$). In contrast, semiconducting MXenes possess very high Seebeck coefficients and relatively good electrical conductivity, resulting in a high power factor. Therefore, the semiconducting MXenes such as
Sc$_2$CT$_2$ (T = F, OH, O) and M$_2$CO$_2$ (M = Ti, Zr, Hf) can be considered as promising candidates for
thermoelectric applications~\cite{M.Khazaei2014_3}.  
Further analyses indicate that the thermal conductivities of MXenes are in the range of 10$-$60 Wm$^{-1}$K$^{-1}$, 
which is in the same range of those for transition metal dichalcogenides~\cite{S.Kumar2016,A.N.Gandi2016,S. Sarikurt2018}. The values of ZT of M$_2$CO$_2$ (M = Ti, Zr, Hf)  
and Sc$_2$CT$_2$ (T = F, OH, O) have been investigated for temperatures of 300$-$900 K: 
The maximum value of ZT for semiconducting MXenes is expected to be less than 1.1~\cite{S.Kumar2016,A.N.Gandi2016,S. Sarikurt2018}.

\subsection{Superconductivity}

Among members of the MAX phase family, Mo$_2$GaC is a superconductor with a critical temperature 
($T_{\rm c}$) of 4 K~\cite{W.Jeitschko1983}.
 2D Mo$_2$C MXene has been exfoliated experimentally from a non-MAX-phase structure, 
Mo$_2$Ga$_2$C~\cite{R.Meshkian2015}, and has already been fabricated with various thicknesses through 
chemical vapor deposition~\cite{C.Xu2015,D.Geng2017}. 
Other alternatives have been suggested to obtain 2D Mo$_2$C in the literature~\cite{W.Sun2016}. 
It has also been observed experimentally that 2D Mo$_2$C is a superconductor 
with $T_{\rm c} <4$~K~\cite{C.Xu2015}.
Theoretically, using Bardeen-Cooper-Schrieffer (BCS) theory~\cite{J.Bardeen1957} 
and DFT electron-phonon interaction 
calculations, the critical temperature $T_{\rm c}$ has been predicted for pristine Mo$_2$C and
Mo$_2$C functionalized with H, OH, O, S, Se, and Br~\cite{J.Lei2017,J.J.Zhang2017}. 
The estimated T$_c$ is in the range from $0$ to $13$~K and the highest $T_{\rm c}$ is around 13 K 
for Mo$_2$C functionalized with H and Br  because of their high electron-phonon coupling. 
The investigation of Fermi surface topology of MAX phases and MXenes with nesting characteristics might be a good strategy to
find new superconductors in this family of materials~\cite{Y.D.Fu2017}.

\subsection{Catalysis}

MXenes with a large surface area, high surface hydrophilicity, and surface activities have attracted applications as catalysts and/or catalyst support.    
As one of the widely studied cases, MXenes have been considered for water electrolysis applications in which H$_2$ and O$_2$ molecules are formed through a hydrogen evolution reaction (HER) at the cathode and an oxygen evolution reaction (OER) at the anode. 
Although platinum is one the most efficient catalysts for water splitting, it has drawbacks due to its high cost and low availability. 
MXenes are superior to other 2D catalysts such as MoS$_2$ and CN because of their higher charge transfer abilities. It has been shown that O-terminated MXenes such as Ti$_2$CO$_2$ and W$_2$CO$_2$ possess Gibbs free energies close to 0 eV for hydrogen adsorption~\cite{C.Ling2016,G.Gao2016_2}, suitable for HER. Experimental evidence of the HER activity of Ti- and Mo-based MXenes suggests that Mo$_2$CT$_x$ and Ti$_3$C$_2$T$_x$ exhibit a good HER activity for generating hydrogen from water and ammonia borane, respectively~\cite{N.Chaudhari2017,J.C.Lei2015}. 
A detailed theoretical investigation, e.g., shown in Fig.~\ref{fig:cata}, indicates that among all MXenes, Mo$_2$CT$_x$ possesses the most promising surface activity for the HER~\cite{Z.Seh2016}. 
The catalytic properties of MXenes can be improved by the introduction of metal atoms such as Fe. This is because less electronegative elements such as Fe can transfer more charge to the O atom of water; hence, the O$-$H bonds of water are weakened, leading to a better HER activity. 
Furthermore, the hybrid structure of Ti$_3$C$_2$/g-C$_3$N$_4$ improves the performance for the OER 
as compared to the case of the pristine MXene because the charge transfer between Ti atoms and graphitic carbon nitride facilitates the electron transfer for O$_2$ evolution~\cite{X.An2018,Y.Sun2018}. 
Another study has shown that Mo$_2$CO$_2$ surfaces can be suitable substrates for Pd atom anchoring to produce O$_2$ from CO molecules~\cite{C.Cheng2018}.

\begin{figure}
\centering
\includegraphics[width=0.85\columnwidth]{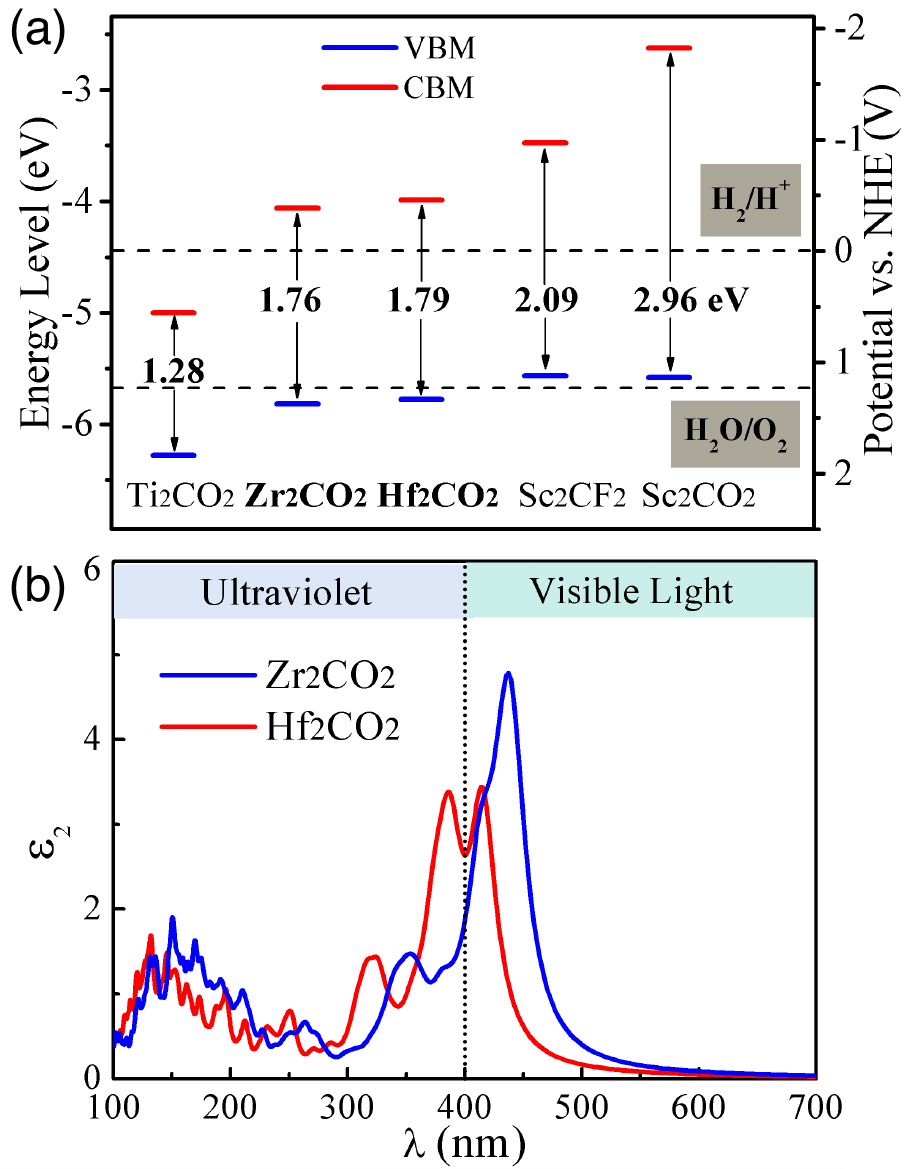}
\caption{(a) Aligned band edges of MXenes relative to the water redox potentials. (b) $\epsilon_2(\omega)$ for O-terminated Zr and Hf MXenes~\cite{Z.Guo2016}.}
\label{fig:photocat}
\end{figure}

\subsection{Photocatalysis}

Some MXenes are semiconducting with light absorption in the visible light region and good catalytic properties. 
Hence, they have 
potential applications for various photocatalytic reactions. 
The production of hydrogen by water splitting on semiconductor substrates via sunlight is an example of such photocatalytic reactions. 
This process involves electron-hole pair generation by absorbing solar light in the photocatalytic material. 
Material with photocatalytic properties  possesses a band gap larger than 1.23 eV for the absorption visible light 
and has redox potentials located between the valence and conduction band edges~\cite{Z.Guo2016}. Owing to semiconducting characteristics, Ti$_2$CO$_2$, Zr$_2$CO$_2$, Hf$_2$CO$_2$, Sc$_2$CO$_2$, and Sc$_2$CF$_2$ can be candidates for photocatalytic applications. 
As shown in Fig.~\ref{fig:photocat}(a), Ti$_2$CO$_2$, Sc$_2$CO$_2$, and Sc$_2$CF$_2$ are not suitable for water splitting because the water redox potentials are beyond the band edges. 
Considering the optical properties of the remaining candidates, Zr$_2$CO$_2$ and Hf$_2$CO$_2$, their absorption peaks are located in the visible light energy window [see Fig.~\ref{fig:photocat}(a)]. 
Therefore, they can be used for the HER via a photocatalytic reaction with water~\cite{Z.Guo2016}.

Further studies indicate that heterostructures of metallic MXenes (e.g., Ti$_3$C$_2$) with other semiconductors (e.g., TiO$_2$) can be used for various photocatalytic reactions. 
For example, the photocatalytic activity of Ti$_3$C$_2$T$_x$ sharply increases with TiO$_2$; an improvement of 400\% is observed for the HER~\cite{A.Shahzad2018}.  In another report, TiO$_2$ nanoplates grown on Ti$_3$C$_2$  shows improved photocatalytic performance, 2.8  times higher than that of pure  TiO$_2$ for CO$_2$ reduction.  This increase in the production can be attributed to the position of the TiO$_2$ conduction band edges, 
which is more negative than the Fermi energy of the MXene; hence, electrons are transferred to Ti$_3$C$_2$ with a smaller energy barrier. Therefore, an electron$-$hole pair can be easily separated without recombination~\cite{J.Low2018}. 
A recent experiment on a Ti-based MXene has successfully prepared CdS/Ti$_3$C$_2$ heterostructure 
and observed an improvement in the photocatalytic activity. The photocatalytic activity of  CdS is low (105 $\mu$mol h$^{-1}$ g$^{-1}$), which varies linearly with the amount of Ti$_3$C$_2$, reaching to a very high activity of 14322 $\mu$mol h$^{-1}$ g$^{-1}$ for the HER.  This improvement can again be correlated with the electron transfer from CdS to the MXene, which pushes the conduction band edge to align with the standard hydrogen electrode (SHE). The versatility of Ti-MXene was verified by fabricating ZnS and Zn doped CdS with Ti$_3$C$_2$ nanoparticles~\cite{J.Ran2017}.

\subsection{Ion batteries}
Owing to their excellent electrical conductivity, large surface areas, and environmentally friendly behavior, 
MXenes have much-attracted attention for applications in energy storage such as capacitors and alkali and metal batteries~\cite{Y.Xie2014_1,M.Naguib2013_2,A.VahidMohammadi2017}. It was found that Ti$_3$C$_2$ monolayers can have a capacity (Li-ion) of 320 mAh g$^{-1}$. Surface functionalization with OH and F groups resulted in the decrease in the specific capacity to 130 and 67 mAh g$^{-1}$, respectively, considering the increase in the atomic weight. 
The calculated diffusion barrier along the minimum energy pathway for a Li atom on a bare Ti$_3$C$_2$ sheet is 
0.005 eV. The introduction of OH and F groups leads to the increase in the diffusion barrier to 0.36 and 1.02 eV, respectively~\cite{Q.Tang2012}. It has also been observed that the formation of additional Li layers on lithiated O-terminated Ti$_3$C$_2$ accounts for the higher gravimetric capacity observed in a delaminated Ti$_3$C$_2$ colloidal solution intercalated with dimethyl sulphoxide~\cite{O.Mashtalir2013}. 
It has also been predicted that the coverage of alkali ions on Ti$_3$C$_2$ and the adsorption energy are related to their effective ionic radii. The increase in the effective ionic radius of an alkali increases the interaction energy while it decreases the coverage~\cite{D.Er2014}. 

The calculations for Ti$_2$C, its functionalized monolayers, and its defective layers indicate that the presence of vacancies affects the diffusion behavior and changes the barrier of lithium atoms~\cite{Q.Wan2018}. 
A systematic analysis of various 2D MXenes finds that Ti$_2$C has a low diffusion barrier for Li in both the pristine and defective state. Li-atom diffusion has a relatively higher energy barrier on a Ti$_2$CX$_2$ 
(X = O, OH)/graphene heterostructure as compared to the Ti$_2$CX$_2$ monolayer and is lower than graphite by 0.5 eV~\cite{Y.Aierken2018}. 
On the other hand, the open-circuit voltage of Ti$_2$CO$_2$/graphene is enhanced to 1.49 V, while sulfur-terminated Ti$_2$C shows a relatively low energy barrier for Li atoms and has the highest affinity for polysulfides~\cite{X.Liu2018}. Moreover, the lithium-sulfur storage capacity of Ti-based MXene was explored recently, 
and it was found that OH groups can entrap the Li$_2$S$_n$ chains formed on top of them MXene~\cite{D.Rao2017}. 

\begin{figure}
\centering
\includegraphics[width=1.0\columnwidth]{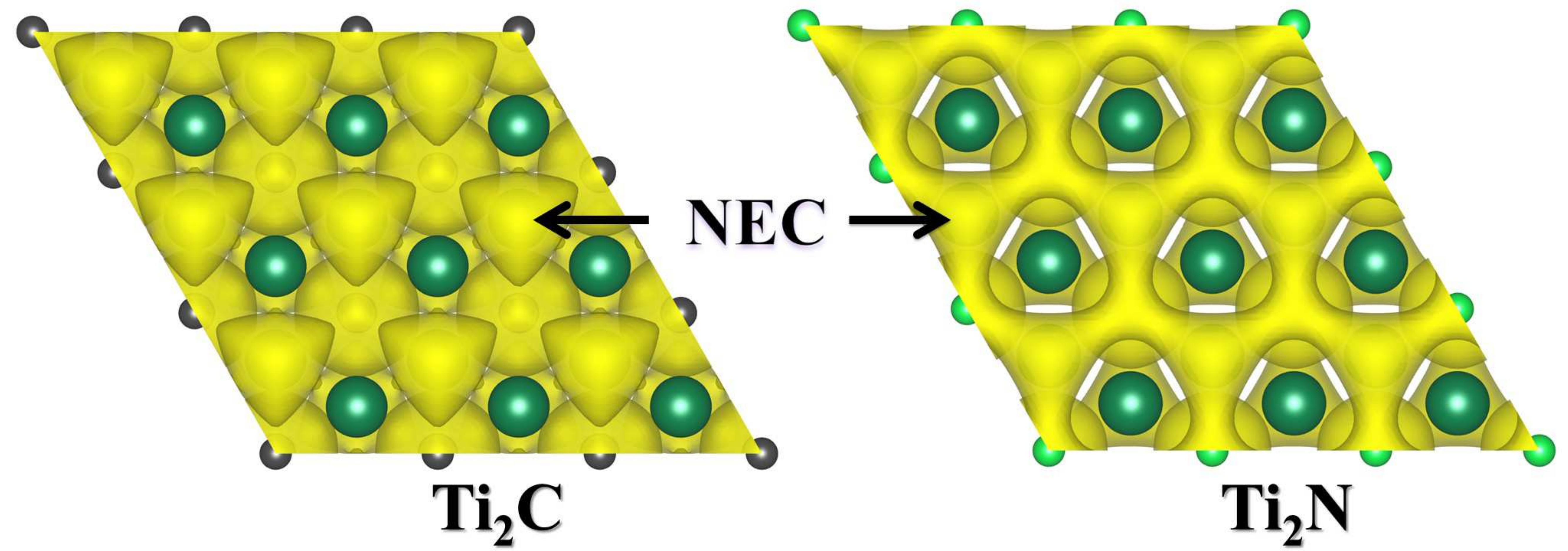}
\caption{Electron localization function isosurfaces (0.5) of Ti$_2$C and Ti$_2$N monolayers. NEC stands for the negative electron cloud.
The dark green, light green, and black spheres indicate Ti, N, and C atoms, respectively~\cite{D.Wang2017}.
}
\label{fig:ion}
\end{figure}

The success achieved in the case of Li-ion batteries has stimulated research on utilizing MXenes for other alkali and alkali-earth metal ion batteries. Accordingly, the adsorption of metal ions such as monovalent (Na and K) and multivalent (Mg, Ca, and Al) ions on layered 
Ti$_2$CX$_2$ (X = O, F, OH) has been studied~\cite{Y.Xie2014_2}. A bare MXene has higher capacities and greater mobilities for metal ions. Na and K ions were found to intercalate more efficiently as compared to other metal ions. Furthermore, O-terminated MXenes decompose into bare MXenes during Mg, Ca, and Al ions intercalation. Monovalent metal ions Na and K have a low diffusion barrier on the Ti$_2$CO$_2$ monolayer as compared to the multivalent metal ions Mg, Ca, and Al.  With a low diffusion barrier for Na on Ti$_3$C$_2$ and its functionalized MXenes, a further increase in the interlayer separation lowers the redox potential. Furthermore, the volume change was observed during sodation and desodation on pristine and F- and O-functionalized Ti-based MXenes, which are favorable for the cycling performance~\cite{Y.X.Yu2016}. 

\begin{figure}
\centering
\includegraphics[width=0.92\columnwidth]{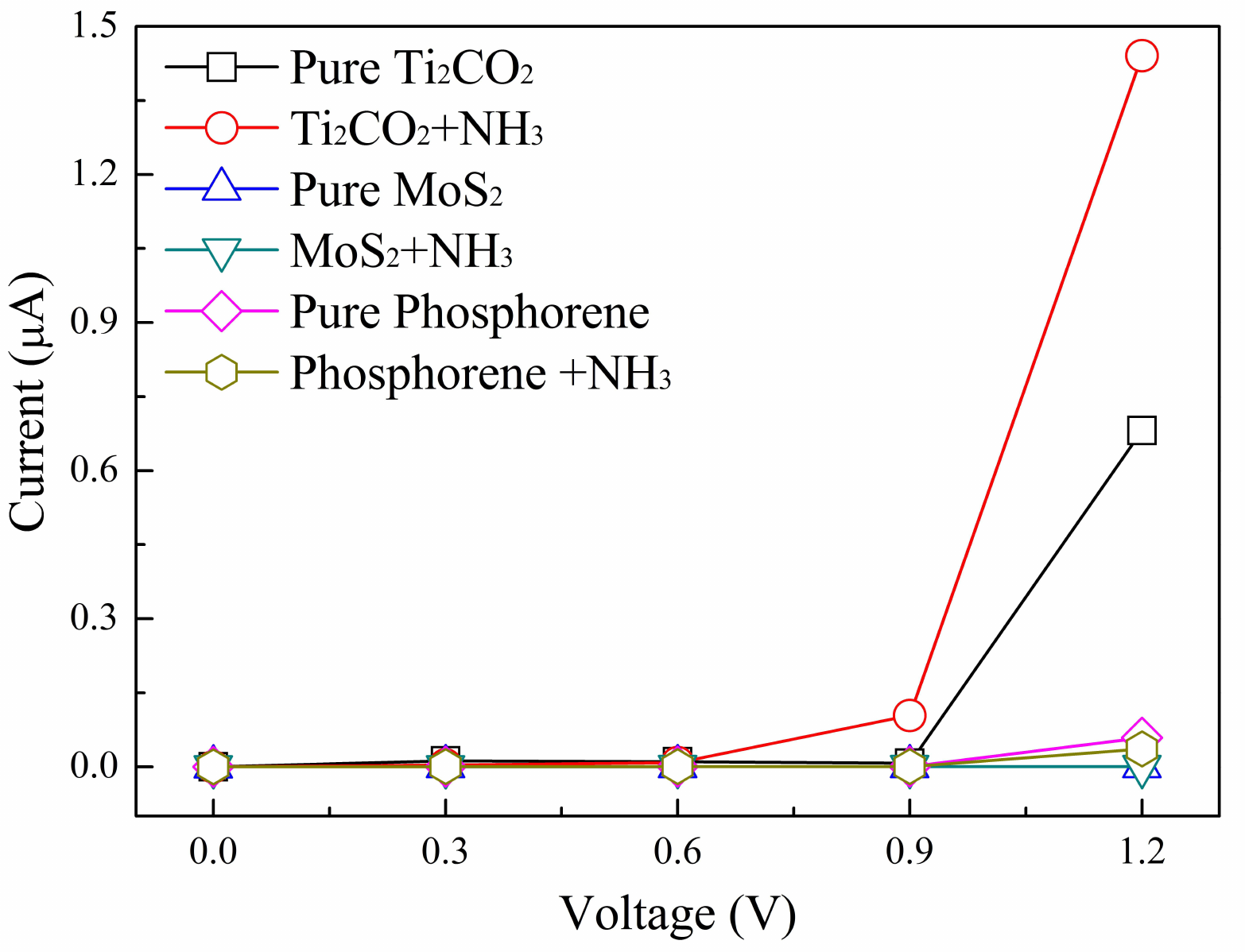}
\caption{Calculated I$-$V curve before and after the adsorption of NH$_3$ on Ti$_2$CO$_2$, MoS$_2$, and phosphorene~\cite{X.F.Yu2015}.
}
\label{fig:sensor}
\end{figure}

With the aim of increasing the electron density, which eventually decreases the diffusion barriers of electron-rich species, 
nitrogen-containing MXenes have been studied. Li adsorption and diffusion on trimetallic Ti$_3$CN and its F-, O-, and OH-functionalized MXenes were studied~\cite{X.Chen2016}. Li atoms preferentially adsorb on the nitrogen side in Ti$_3$CN. In the functionalized materials, Li atoms tend to be adsorbed on the carbon side. The diffusion barriers are in the range of 0.2$-$0.3 eV for functionalized Ti$_3$CN. The potential use of Ti$_2$N and its functionalized derivatives for use as anode materials was recently tested~\cite{D.Wang2017}. The adsorption energies of cations including Li, Na, K, Mg, Ca, and Al and the open-circuit voltage suggest that Ti$_2$N monolayers are a promising candidate for the anode material in rechargeable batteries. 
An electron localization function analysis of carbide and nitride MXenes suggests that the N atom in Ti$_2$N needs one less electron from Ti to form an ionic bond, thereby creating a nonbonding electron cloud on the Ti$_2$N surface. 
The presence of negative electron cloud results in a low diffusion barrier for the cations (see Fig.~\ref{fig:ion}). 
Owing to their rich electronic nature, many attempts have been carried out to use other transition metal MXenes for battery applications, e.g., V$_{n+1}$C$_n$, Nb$_2$C, Cr$_2$CO$_2$, and Mo$_2$C~\cite{J.Hu2014,L.Bai2016,D.Sun2016,J.Zhu2015,L.Bai2018,F.Li2016,Z.Zou2018,Q.Sun2016}.

\begin{figure}
\centering
\includegraphics[width=0.99\columnwidth]{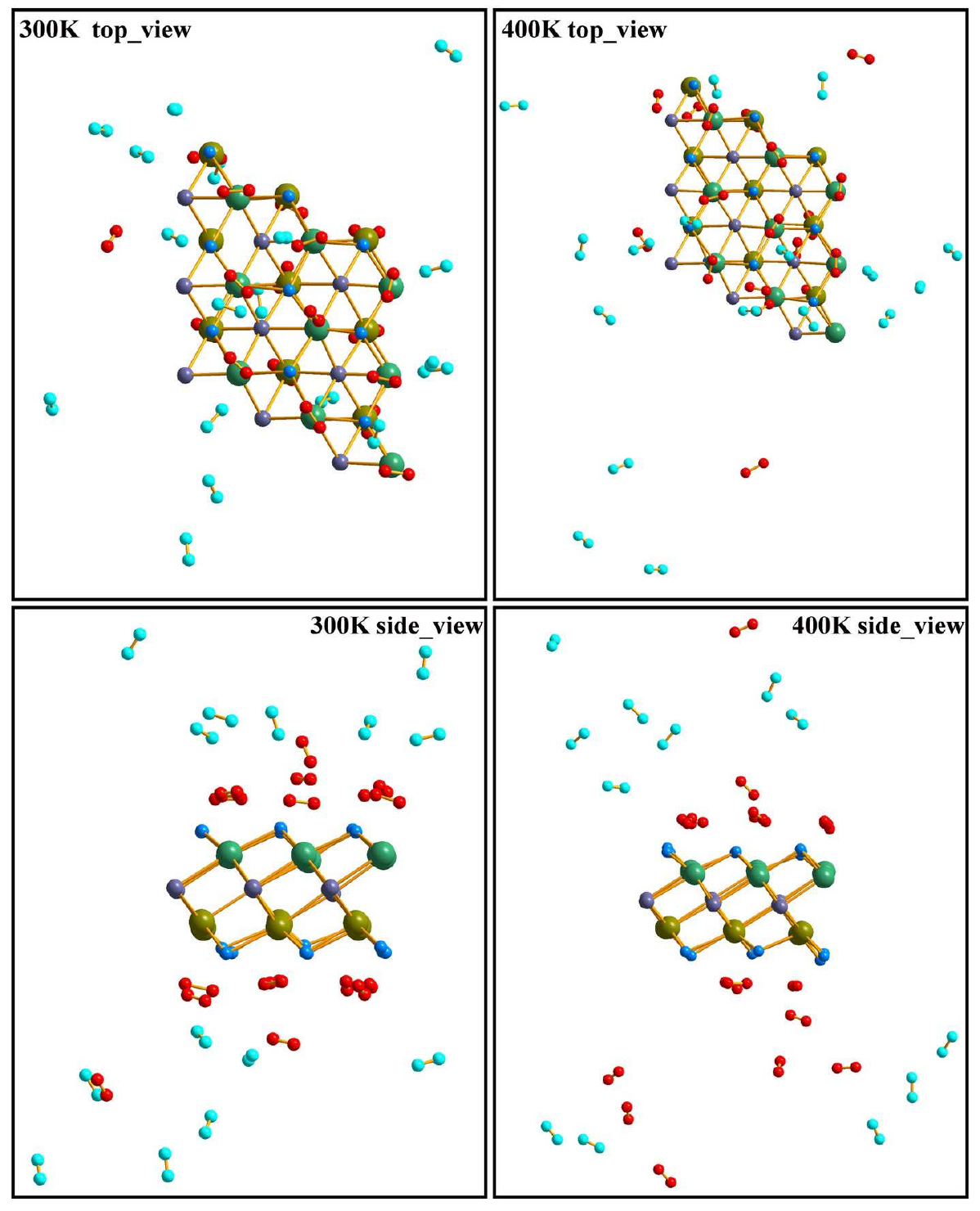}
\caption{Snapshots of molecular dynamics for hydrogen adsorption on Ti$_2$C at 300 and 400 K after 1.5 ps 
viewed from top and side of Ti$_2$C.
Physisorbed and adsorbed H$_2$ by Kubas interactions are indicated by the cyan and red spheres, 
respectively~\cite{Q.Hu2013}.}
\label{fig:storage}
\end{figure}

\subsection{Gas sensors}

Recently, layered materials such as metal oxides and graphene have been extensively studied as gas sensors. 
MXenes have also been shown to be potential gas sensors due to the direct charge transfer 
mechanism~\cite{M.Khazaei2016_1}.  Recent experiments have observed that 
T$_3$C$_2$ exhibits affinity towards CO$_2$ capture selectivity over N$_2$~\cite{I.Persson2018_2}. 
Furthermore, T$_3$C$_2$ sensors can successfully detect gasses such as  ethanol, methanol,
acetone, and ammonia at room temperature~\cite{E.Lee2017}.
First-principles calculations have found 
that gas molecules such as H$_2$, CO$_2$, O$_2$, N$_2$, NO$_2$, CH$_4$, and CO are subject to 
physisorption on Ti$_2$C, while NH$_3$ undergoes chemisorption~\cite{X.F.Yu2015}. 
The adsorption of NH$_3$ molecules increases the 
electrical conductivity (see Fig.~\ref{fig:sensor}), leading to the high sensitivity of Ti$_2$CO$_2$ to NH$_3$ 
as compared to MoS$_2$ and phosphorene. Moreover, it has been shown theoretically 
that the oxygen functionalized 
MXenes exhibit ehancement of the selectivity~\cite{A.Junkaew2018}. For example, 
Mo$_2$CO$_2$ and V$_2$CO$_2$ have selectivity towards NO; and Nb$_2$CO$_2$ and Ti$_2$CO$_2$ 
shows selectivity toward NH$_3$~\cite{A.Junkaew2018}. 
Very recently, it has been predicted that a Sc$_2$CO$_2$ monolayer is a putative candidate for SO$_2$ gas sensing~\cite{S.Ma2017}. Among the MXenes studied, SO$_2$ has a 
high adsorption strength and higher charge transfer towards Sc$_2$CO$_2$. By applying tensile strains or external electric fields, the adsorption strength of SO$_2$ can be controlled, and even the reversible release of captured SO$_2$ can be achieved~\cite{S.Ma2017}.

\subsection{Hydrogen storage}
MXenes have also been investigated for possible hydrogen storage. For example, it was predicted that hydrogen molecules on the surfaces of Ti$_2$C are adsorbed in molecular form and then dissociated to form hydrides, while hydrogen molecules in the second and third layers from the surfaces are adsorbed in molecular form~\cite{Q.Hu2013}. The hydrides constitute  1.7 wt\%, and molecular hydrogen has a reversible capacity of 3.4 wt\%. Molecular dynamics studies show that the hydrogen that exists with a Kubas-type interaction departs, leaving the hydrides intact on the surface (see Fig.~\ref{fig:storage}). This clearly justifies the reversible uptake of hydrogen on the Ti$_2$C layer. A similar conclusion has been reached for Sc$_2$C and V$_2$C~\cite{Q.Hu2013, Q.Hu2014}. A recent study has shown that the hydrogen storage capacity of Cr$_2$C can reach 7.6 wt\%, where 1.2 wt\% is due to chemisorption, 3.2 wt\% is due to the Kubas interaction, and 3.2 wt\% exists in molecular form with a weak electrostatic interaction for hydrogen storage~\cite{A.Yadav2016}. Hence, a reversible hydrogen storage capacity of 6.4 wt\% can be achieved with a binding energy 0.1$-$0.44 eV/H$_2$ on the Cr$_2$C surface~\cite{A.Yadav2016}.

\section{Conclusions and outlook}
MXenes with rich physics and chemistry, as well as deep structural and compositional diversities, offer a fertile ground for performing fundamental and multidisciplinary research. Nevertheless, until now the synthesis of MXenes with a uniform and pure surface termination is an experimental challenge, and it is an obstacle in realizing the full potential of proposed many electronic-based phenomena and applications.  Therefore, it is valuable to investigate the mixed functionalization of MXene to explore its applications that are not significantly affected by surface termination. The mixed functionalization can also lead to the formation of various interfaces, such as sharp and diffusive interfaces, at the surface of MXenes. Experimental attempts should be made to isolate the pristine MXenes for their full characterization and subsequently controlled functionalization as well. 

Different types of magnetic orders with the large SOC in MXenes promise novel quantum anomalous Hall systems. The response to the external electric field, mechanical strain, and photo-irradiation are the key to tailoring the novel phenomena and applications of MXene-based devices.  Exploring the effect of the gate voltage on controlling the magnetic orders in MXenes is an exciting area of science. Moreover, very little has been explored about the superconductivity in MXenes.  In this regard, the theoretical investigation of the Fermi surface nesting and electron-phonon couplings in various MAX phases and MXenes are crucial.

Most probably, vertical and lateral heterostructures of MXene with other 2D materials will be synthesized in the near future. The effect of van der Waals interactions on stability  and electronic properties of multilayered MXenes should be investigated. The interfacial physics of such junctions will be of fundamental interest for device applications. Furthermore, the effect of electric field on monolayers have already been investigated. This should be extended to few layers, multilayers, and heterostructures of MXenes to examine the electron or hole accumulation on the surface and interface and their influence on electronic properties.  The effect of electron/hole injection on chemical bonding and elastic properties of MXene should be investigated for making artificial muscles and various actuator applications. The electron-hole coupling of the heterostructure of MXenes should be examined for solar cell applications. The electronic properties of MXenes in the presence of the magnetic field and the absorption/reflection of the magnetic field 
are vital for understanding shielding effect.

The development of classical potentials or phase field parameters for performing large-scale molecular dynamics simulation is necessary to study the biological and composite applications of MXenes,  
to simulate the chemical exfoliation of MAX phases using different etchant solutions, 
to understand the mechanism of hydration of different charged alkali metals between MXene layers and its effects on interlayer distances,  
and to study the hydrophilicity of MXene surfaces. 
The effects of using a mixture of A elements in synthesis and exfoliation of MAX phases should be addressed computationally. 
It is also interesting to investigate the mechanical exfoliation of MAX phases both theoretically and experimentally 
in detail and compare the resulting structures with those obtained from chemical exfoliations.

It is highly desirable to study the thermal conductivity of MXenes and screen out the promising thermoelectric candidates.  The estimation of relaxation time is essential to evaluate the electrical conductivity and figure of merits of MXenes and MAX phases. The effect of compression on thermoelectric properties of MXenes is of fundamental interest. The investigation of electron transport and optical conductivity of MXenes is essential for electronic device applications. 
It is also worth determining which of MXenes can be good candidates for cathode or anode electrode applications. Developing new in- and out-of-plane ordered double transition metals MXenes is also important for assessing various magnetic orderes. The physics behind the excellent piezoelectricity of the in-plane ordered double transition metals should also be investigated.  The possible Rashba effect in janus MXenes needs to be examined in more details.

The effects of M and/or X vacancies in MXenes~\cite{X.Sang2016}  should be explored on various applications 
such as ion batteries and electrode structural flexibility. The potential applications of vacancy-ordered in-plane MXenes for gas storage should also be studied.  
 Detailed and systematic analyses are required on nanoribbons~\cite{S.Zhao2015}, nanoscrolls, clusters, or quantum dots~\cite{Q.Xu2018} of MXenes.  Finally, it is of great interest to synthesize the multifunctional group III based MXenes (e.g., Sc$_2$CO$_2$), which, due to their surface polarization, might be good candidates for polarization-driven catalysis~\cite{A.Kakekhani2016}.

\section*{Acknowledgments}
M.K. is grateful to RIKEN Advanced 
Center for Computing and Communication (RCCC) for the allocation of the computational resources of the RIKEN 
supercomputer system (HOKUSAI GreatWave). Some of the calculations were also performed on the Numerical 
Materials Simulator at NIMS. M.K. gratefully acknowledges the support by a Grant-in-Aid for Scientific Research 
(No. 17K14804) from MEXT Japan. 
A.M. and A.K.S. acknowledge the computation facilities at the Materials Research Centre, 
the Thematic Unit of Excellence, and the Supercomputer Education and
Research, Indian Institute of Science, Bangalore, India.  A.K.S. and A.M. are thankful for support from DST Nano Mission, and A.M. acknowledges UGC India for a Senior Research Fellowship.

\section*{References}



\begin{thebibliography}{00}

 


\bibitem{K.S.Novoselov2016}
K. S. Novoselov, A. Mishchenko, A. Carvalho, A. H. Castro Neto,  
2D materials and van der Waals heterostructures,
Science 353 (2016) aac9439.

\bibitem{M.Naguib2011}
M. Naguib, M. Kurtoglu, V. Presser, J. Lu, J. Niu, M. Heon, L. Hultman, Y. Gogotsi, M. W. Barsoum,  
Two-dimensional nanocrystals produced by exfoliation of Ti$_3$AlC$_2$.
Adv. Mater. 23 (2011) 4248$-$4253.

\bibitem{M.Naguib2012}
 M. Naguib, O. Mashtalir, J. Carle, V. Presser, J. Lu, L. Hultman, Y. Gogotsi, M. W. Barsoum, 
Two-dimensional transition metal carbides,
ACS Nano 6 (2012) 1322$-$1331.

\bibitem{M.W.Barsoum2000}
M. W. Barsoum, 
The M$_{N+1}$AX$_N$ phases: A new class of solids: Thermodynamically stable nanolaminates,
Prog. Solid State Chem. 28 (2000) 201$-$281.



\bibitem{M.A.Hope2016}
M. A. Hope, A. C. Forse, K. J. Griffith, M. R. Lukatskaya, M. Ghidiu, Y. Gogotsi, C. P. Grey,
NMR reveals the surface functionalisation of Ti$_3$C$_2$ MXene, 
Phys. Chem. Chem. Phys. 18 (2016) 5099.

\bibitem{J.Halim2016}
J. Halim, K. M. Cook, M. Naguib, P. Eklund, Y. Gogotsi, J. Rosen, and M. W. Barsoum, 
X-ray photoelectron spectroscopy of select multi-layered transition metal carbides (MXenes). 
Appl. Surf. Sci. 362 (2016) 406$-$417.

\bibitem{M.Ghidiu2014_2}
M. Ghidiu, M. R. Lukatskaya, M. Q. Zhao, Y. Gogotsi, and M. W. Barsoum, 
Conductive two-dimensional titanium carbide `clay' with high volumetric capacitance,
Nature 516 (2014) (7529) 78$-$81.

\bibitem{J.Halim2014}
J. Halim, M. R. Lukatskaya, K. M. Cook, J. Lu, C. R. Smith, L.-\AA ke N\"{a}slund, S. J. May, L. Hultman, Y. Gogotsi, P. Eklund, and M. W. Barsoum,
Transparent conductive two-dimensional titanium carbide epitaxial thin films,
Chem. Mater. 26 (2014) 2374$-$2381.


\bibitem{M.W.Barsoum2011}
M. W. Barsoum, M. Radovic,
Elastic and Mechanical Properties of the MAX Phases,
Annu. Rev. Mater. Res. 41 (2011) 195$-$227.

\bibitem{M.Khazaei2014_1} 
M. Khazaei, M. Arai, T. Sasaki, M. Estili, Y. Sakka, 
Trends in electronic structures and structural properties of MAX phases: a first-principles study on 
M$_2$AlC (M = Sc, Ti, Cr, Zr, Nb, Mo, Hf, or Ta), M$_2$AlN, and hypothetical M$_2$AlB phases
J. Phys.: Condens. Matter 26 (2014) 505503.

\bibitem{M.Khazaei2017}
 M. Khazaei, A. Ranjbar, M. Arai, T. Sasaki, S. Yunoki,  
Electronic properties and applications of MXenes: a theoretical review,
J. Mater. Chem. C 5 (2017) 2488$-$2503.

\bibitem{N.M.Caffrey2018}
N. M. Caffrey,
Effect of mixed surface terminations on the structural and 
electrochemical properties of two-dimensional Ti$_3$C$_2$T$_2$ and V$_2$CT$_2$ MXenes multilayers, 
Nanoscale 10 (2018) 13520$-$13530.

\bibitem{T.Hu2018}
T. Hu, M. Hu, B. Gao, W. Li, X. Wang,
Screening surface structure of MXenes by high-throughput
computation and vibrational spectroscopic confirmation,
J. Phys. Chem. 122 (2018) 18501$-$18509.

\bibitem{I.Persson2018}
I. Persson, L.-\AA~N\"{a}slund, J. Halim, M. W. Barsoum, V. Darakchieva, J. Palisaitis, J. Rosen, P. O. \AA Persson,
On the organization and thermal behavior of functional groups on Ti$_3$C$_2$ MXene surfaces in vacuum,
2D Mater. 5 (2018) 015002.

\bibitem{B.Anasori2017_1}
 B. Anasori, M. R. Lukatskaya, Y. Gogotsi,
 2D metal carbides and nitrides (MXenes) for energy storage,
Nat. Rev. 2 (2017) 16098.

\bibitem{B.M.Jun2018}
B.-M. Jun, S. Kim, J. Heo, C. M. Park, N. Her, M. Jang, Y. Huang, J. Han, Y. Yoon,
Review of MXenes as new nanomaterials for energy storage/delivery,
and selected environmental applications.
Nano Res. (2018); https://doi.org/10.1007/s12274-018-2225-3. 

\bibitem{J.Pang2019}
 J. Pang,  R. G. Mendes,  A. Bachmatiuk, L. Zhao,  H. Q. Ta,  T. Gemming, H. Liu,  Z. Liu, Mark H. Rummeli,
Applications of 2D MXenes in energy conversion and storage systems,
Chem. Rev. Soc. 48 (2019) 72$-$133.

\bibitem{A.L.Ivanovskii2013}
A. L. Ivanovskii and A. N. Enyashin,
Graphene-like transition-metal nanocarbides and nanonitrides.
Russ. Chem. Rev. 82 (2013) 735$-$746. 

\bibitem{N.K.Chaudhari2017}
 N. K. Chaudhari,  H. Jin,  B. Kim,  D. S. Baek, S. H. Joo, K. Lee,
MXene: an emerging two-dimensional material for future energy conversion and storage applications, 
J. Mater. Chem. A 5 (2017) 24564$-$24579.

\bibitem{J.Zhu2017}
J. Zhu, E. Ha, G. Zhao, Y. Zhou, D. Huang, G. Yue, L. Hu, N. Sun, Y. Wang, L. Y. S. Lee, C. Xu, K.-Y. Wong, D. Astruc, and P. Zhao, 
Recent advance in MXenes: A promising 2D material for catalysis, sensor and chemical adsorption,
Coord. Chem. Rev. 352 (2017) 306$-$327. 

\bibitem{H.Wang2018} 
H. Wang, Y. Wu, X. Yuan, G. Zeng, J. Zhou, X. Wang, J. W. Chew,
Clay-inspired MXene-based electrochemical devices and photo-electrocatalyst: state-of-the-art progresses and challenges,
Adv. Mater. 30 (2018) 1704561. 

\bibitem{X.Li2018} 
X. Li, C. Wang, Y. Cao, G. Wang,
Functional MXene materials: progress of their applications,
Chem. Asian J. 13 (2018) 2742$-$2757.

\bibitem{X.Zhang2018}
X. Zhang, Z. Zhang, Z. Zhou,
MXene-based materials for electrochemical energy storage,
J. Energy Chem. 27 (2018) 73$-$85.

\bibitem{Y.Zhang2018}
 Y. Zhang,  L. Wang,  N. Zhang,  Z. Zhou,
Adsorptive environmental applications of MXene nanomaterials: a review,
 RSC Adv. 8 (2018) 19895$-$19905.

\bibitem{K.Hantanasirisakul2018} 
 K. Hantanasirisakul and Y. Gogotsi,
 Electronic and optical properties of 2D transition metal carbides and nitrides (MXenes)
 Adv. Mater. (2018), 1804779. 
 
 \bibitem{H.Lin2018}
 H. Lin, Y. Chen, and J. Shi,
 Insights into 2D MXenes for versatile biomedical 
applications: current advances and challenges ahead,
Adv. Sci. 5 (2018) 1800518.



\bibitem{M.Naguib2013}
 M. Naguib, J. Halim, J. Lu, K. M. Cook, L. Hultman, Y. Gogotsi, M. W. Barsoum,  
 New Two-dimensional niobium and vanadium carbides as promising materials for Li-Ion batteries,
J. Am. Chem. Soc. 135 (2013) 15966$-$15969.

\bibitem{M.Ghidiu2014}
M. Ghidiu, M. Naguib, C. Shi, O. Mashtalir, L. M. Pan, B. Zhang, J. Yang, Y. Gogotsi, S. J. L. Billinge, M. W. Barsoum,
Synthesis and characterization of two-dimensional Nb$_4$C$_3$ (MXene),
Chem. Commun. 50 (2014) 9517$-$9520.

\bibitem{J.Zhou2016}
 J. Zhou, X. Zha, F. Y. Chen, Q. Ye, P. Eklund, S. Du, Q. Huang,  
 A two-dimensional zirconium carbide by selective etching of Al$_3$C$_3$ from nanolaminated Zr$_3$Al$_3$C$_5$
Angew. Chem.-Int. Edit. 55 (2016) 5008$-$5013.

\bibitem{P.Urbankowski2016}
 P. Urbankowski, B. Anasori, T. Makaryan, D. Er, S. Kota, P. L. Walsh, M. Zhao, V. B. Shenoy, M. W. Barsoum, Y. Gogotsi,
Synthesis of two-dimensional titanium nitride Ti$_4$N$_3$ (MXene),
Nanoscale 8 (2016) 11385$-$11391.


\bibitem{B.Soundiraraju2017}
 B. Soundiraraju, B. K. George, 
 Two-dimensional titanium nitride (Ti$_2$N) MXene: synthesis, characterization, and potential application as surface-enhanced Raman scattering substrate,
 ACS Nano 11 (2017) 8892$-$8900.

\bibitem{M.Alhabeb2018}
M. Alhabeb, K. Maleski, T. S. Mathis, A. Sarycheva, C. B. Hatter, S. Uzun,
A. Levitt, and Y. Gogotsi,
Selective etching of silicon from Ti$_3$SiC$_2$ (MAX) To Obtain 2D Titanium Carbide (MXene),
Angew. Chem. Int. Ed. Engl. 57 (2018) 5444$-$5448.

\bibitem{M.H.Tran2018}
M. H. Tran, T. Sch\"{a}fer, A. Shahraei, M. D\"{u}rrschnabel, L. Molina-Luna, U. I. Kramm, C. S. Birkel,
Adding a new member to the MXene family: Synthesis, structure, and electrocatalytic activity for the hydrogen evolution reaction of V$_4$C$_3$T$_x$,
ACS Appl. Energy Mater. 1 (2018) 3908$-$3914.

\bibitem{J.Yang2016_2} 
J. Yang, M. Naguib, M. Ghidiu, L.-M. Pan, J. Gu, J. Nanda, J. Halim, Y. Gogotsi, M. W. Barsoum, 
Two-dimensional Nb-based M$_4$C$_3$ solid solutions (MXenes),
J. Am. Ceramic Soc. 99 (2016) 660$-$666.


\bibitem{B.Anasori2015_1}
B. Anasori, M. Dahlqvist, J. Halim, E. J. Moon, J. Lu, B. C. Hosler, E. N. Caspi, S. J. May, L. Hultman, P. Eklund, J. Ros\'{e}n, M. W. Barsoum,  
Experimental and theoretical characterization of ordered MAX phases Mo$_2$TiAlC$_2$ and Mo$_2$Ti$_2$AlC$_3$,
J. Appl. Phys. 118 (2015) 94304.

\bibitem{B.Anasori2015_2}
 B. Anasori, Y. Xie, M. Beidaghi, J. Lu, B. C. Hosler, L. Hultman, P. R. C. Kent, Y. Gogotsi, M. W. Barsoum, 
Two-dimensional, ordered, double transition metals carbides (MXenes),
ACS Nano 9 (2015) 9507$-$9516.

\bibitem{R.Meshkian2017}
R. Meshkian, Q. Tao, M. Dahlqvist, J. Lu, L. Hultman and J. Rosen, 
Theoretical stability and materials synthesis of a chemically ordered MAX phase, Mo$_2$ScAlC$_2$, and its two-dimensional derivate Mo$_2$ScC$_2$ MXene,
Acta Mater. 125 (2017) 476$-$480.



\bibitem{Q.Tao2017}
Q. Tao, M. Dahlqvist, J. Lu, S. Kota, R. Meshkian, J. Halim, J. Palisaitis, L. Hultman, M. W. Barsoum, P. O. Persson, J. Rosen, 
Two-dimensional Mo$_1.33$C MXene with divacancy ordering prepared from parent 3D laminate with in-plane chemical ordering,
Nat. Commun. 8 (2017) 14949.

\bibitem{M.Dahlqvist2017}
M. Dahlqvist, J. Lu, R. Meshkian, Q. Tao, L. Hultman, J. Rosen,
Prediction and synthesis of a family of atomic laminate phases with Kagom\'{e}-like and in-plane chemical ordering,
Sci. Adv. 3 (2017) e1700642.

\bibitem{J. Lu2017}
J. Lu, A. Thore, R. Meshkian, Q. Tao, L. Hultman, J. Rosen, 
Theoretical and experimental exploration of a novel in-plane chemically-ordered,
(Cr$_{2/3}$M$_{1/3}$)$_2$AlC i-MAX Phase with M=Sc and Y, 
Cryst. Growth Des. 17 (2017) 5704$-$5711.

\bibitem{R.Meshkian2018}
R. Meshkian, M. Dahlqvist, J. Lu, B. Wickman, J. Halim, 
J. Th\"{o}rnberg, Q. Tao, S. Li, S. Intikhab, J. Snyder, M. W. Barsoum, M. Yildizhan, J. Paiisaitis, L. Hultman, P. O. \AA. Persson, J. Rosen,
W-based atomic laminates and their 2D derivative W$_{1.33}$C MXene with vacancy ordering,
Adv. Mater. 30 (2018) 1706409.

\bibitem{L.Chen2018}
L. Chen, Martin Dahlqvist, T. Lapauw, B. Tunca, F. Wang, J. Lu, R. Meshkian, K. Lambrinou, B. Blanpain, J. Vleugels, J. Rosen
Inorg. Chem. 57 (2018) 6237$-$6244.

\bibitem{M.Dahlqvist2018}
M. Dahlqvist, A. Petruhins, J. Lu, L. Hultman, J. Rosen,
Origin of chemically ordered atomic laminates (i-MAX): expanding the elemental space by a theoretical/experimental approach,
ACS Nano 12 (2018) 7761$-$7770.

\bibitem{J.Halim2018}
J. Halim, J. Palisaitis, J. Lu, J. Th\"{o}rnberg, E. J. Moon, M. Precner, P. Eklund, P. O. \AA. Persson, M. W. Barsoum, J. Rosen, 
Synthesis of two-dimensional Nb$_{1.33}$C (MXene) with randomly distributed vacancies by etching of the quaternary 
solid solution (Nb$_{2/3}$Sc$_{1/3}$)$_2$AlC MAX Phase, 
ACS Appl. Nano Mater. 6 (2018) 2455$-$2460.











\bibitem{P.Srivastava2016}
P. Srivastava, A. Mishra, H. Mizuseki, K. R. Lee, A. K. Singh,
Mechanistic insight into the chemical exfoliation and functionalization of Ti$_3$C$_2$ MXene, 
ACS Appl. Mater. Interfaces 8 (2016) 24256$-$24264.



\bibitem{M.Khazaei2018_1}
M. Khazaei, A. Ranjbar, K. Esfarjani, D. Bogdanovski, R. Dronskowski, S. Yunoki,
Insights into exfoliation possibility of MAX phases to MXenes,
Phys. Chem. Chem. Phys. 20 (2018) 8579$-$8592.

\bibitem{Z.Guo2015}
Z. Guo, L. Zhu, J. Zhou, Z. Sun, 
Microscopic origin of MXenes derived from layered MAX phases,
RSC Adv. 5 (2015) 25403$-$25408.


\bibitem{M.Khazaei2014_2} 
M. Khazaei, M. Arai, T. Sasaki, M. Estili, Y. Sakka, 
The effect of the interlayer element on the exfoliation of layered Mo$_2$AC (A= Al, Si, P, Ga, Ge, As or In) 
MAX phases into two-dimensional Mo$_2$C nanosheets,
Sci. Tech. Adv. Mater. 15 (2014) 014208.



\bibitem{M.Yi2015}
M. Yi, Z. Shen,
A review on mechanical exfoliation for the scalable production of graphene,
J. Mater. Chem. A 3 (2015) 11700. 

\bibitem{A.Mishra2017}
A. Mishra, P. Srivastava, A. Carreras, I. Tanaka, H. Mizuseki, K. R. Lee, A. K. Singh,
Atomistic origin of phase stability in oxygen-functionalized MXene: a comparative study,
J. Phys. Chem. C 121 (2017) 18947$-$18953.

\bibitem{M.Khazaei2013}
M. Khazaei, M. Arai, T. Sasaki, C.-Y. Chung , N. S. Venkataramanan, M. Estili, Y. Sakka, Y. Kawazoe, 
Novel electronic and magnetic properties of two-dimensional transition metal carbides and nitrides,
Adv. Funct. Mater. 23 (2013) 2185$-$2192.




\bibitem{U.Yorulmaz2016}
U. Yorulmaz, A. \"{O}zden, N. K. Perkg\"{o}z, F. Ay, C. Sevik,
Vibrational and mechanical properties of single layer MXene structures: a first-principles investigation,
Nanotechnology 27 (2016) 335702.

\bibitem{M.Khazaei2014_3}
M. Khazaei, M. Arai, T. Sasaki, M. Estili, Y. Sakka,
Two-dimensional molybdenum carbides: potential thermoelectric materials of the MXene family,
Phys. Chem. Chem. Phys. 16 (2014) 7841$-$7849.


  \bibitem{Y.Xie2014_1}
Y. Xie, M. Naguib, Y. N. Mochalin, M. W. Barsoum, Y. Gogotsi, X. Yu, K.-W. Nam, X.-Q. Yang, A. I. Kolesnikov, P. R. C. Kent, 
 Role of surface structure on Li-ion energy storage capacity of two-dimensional transition-metal carbides, 
 J. Am. Chem. Soc. 136 (2014) 6385$-$6394.



\bibitem{T.Hu2017}
T. Hu, Z. Li, M. Hu, J. Wang, Q. Hu, Q. Li, X. Wang,
Chemical origin of termination-functionalized MXenes: Ti$_3$C$_2$T$_2$ as a case study,
J. Phys. Chem. C 121 (2017) 19254$-$19261.





\bibitem{M.Ashton2016_2}
M. Ashton, K. Mathew, R. G. Hennig, S. B. Sinnott,
Predicted Surface Composition and Thermodynamic Stability of MXenes in Solution,
J. Phys. Chem. C 120 (2016) 3550.

\bibitem{M.Naguib2014}
M. Naguib, O. Mashtalir, M. R. Lukatskaya, B. Dyatkin, C. Zhang, V. Presser, Y. Gogotsi, M. W. Barsoum,
One-step synthesis of nanocrystalline transition metal oxides on thin sheets of disordered graphitic carbon by oxidation of MXenes, 
Chem. Commu. 50 (2014) 7420. 



\bibitem{H.Weng2015_1}
 H. Weng, A. Ranjbar, Y. Liang, Z. Song, M. Khazaei, S. Yunoki, M. Arai, Y. Kawazoe, Z. Fang, X. Dai,
 Large-gap two-dimensional topological insulator in oxygen functionalized MXene, 
Phys. Rev. B 92 (2015) 075436.

\bibitem{M.Khazaei2016_2}
M. Khazaei, A. Ranjbar, M. Arai, S. Yunoki, 
Topological insulators in the ordered double transition metals M$'_2$M$''$C$_2$ MXenes (M$'$=Mo, W; M$''$=Ti, Zr, Hf),
Phys. Rev. B 94 (2016) 125152.


\bibitem{L.Li2016}
L. Li, 
Lattice dynamics and electronic structures of Ti$_3$C$_2$O$_2$ and Mo$_2$TiC$_2$O$_2$ (MXenes): The effect of Mo substitution,
Comput. Mater. Sci. 124 (2016) 8$-$14.
 

\bibitem{C.Si2016}
C. Si, K.-H. Jin, J. Zhou, Z. Sun, F. Liu,
Large-Gap Quantum Spin Hall State in MXenes: d-Band Topological Order in a Triangular Lattice,
Nano Lett. 16 (2016) 6584$-$6591.

\bibitem{C.Si2016_2}
C. Si, J. You, W. Shi, J. Zhou, Z. Sun,
Quantum spin Hall phase in Mo$_2$M$_2$C$_3$O$_2$ (M = Ti, Zr, Hf) MXenes 
J. Mater. Chem. C 4 (2016) 11524$-$11529.

\bibitem{Y.Liang2017}
Y. Liang, M. Khazaei, A. Ranjbar, M. Arai, S. Yunoki, Y. Kawazoe, H. Weng, Z. Fang,
Theoretical prediction of two-dimensional functionalized MXene nitrides as topological insulators,
Phys. Rev. B 96 (2017) 195414.

\bibitem{M.Z.Hasan2010}
M. Z. Hasan, C. L. Kane,  
Colloquium: Topological insulators,
Rev. Mod. Phys. 82 (2010) 3045.

\bibitem{E.Balci2018}
E. Balci, \"{U}. \"{O}. Akkus, S. Berber,
Controlling topological electronic structure of multifunctional MXene layer,
Appl. Phys. Lett. 113 (2018) 083107.


\bibitem{M.Khazaei2018_2}
M. Khazaei, V. Wang, C. Sevik, A. Ranjbar, M. Arai, S. Yunoki, 
Electronic structures of iMAX phases and their two-dimensional derivatives: A family of piezoelectric materials, 
Phys. Rev. Materials 2 (2018) 074002.


\bibitem{anant}
aNANt: a functional materials database, http://anant.mrc.iisc.ac.in/.

\bibitem{A.C.Rajan2018}
A. C. Rajan, A. Mishra, S. Satsangi, R. Vaish, H. Mizuseki, K.-R. Lee, A. K. Singh,
Machine-learning assisted accurate band gap predictions of functionalized MXene,
Chem. Mat. 30 (2018) 4031$-$4038.


\bibitem{X.F.Yu2015_2}
X.-F. Yu, Jian -B. Cheng, Z.-B. Liu, Q.-Z. Li, W.-Z. Li, X. Yang, and B. Xiao,
The band gap modulation of monolayer Ti$_2$CO$_2$ by strain,
RSC Adv. 5 (2015) 30438. 

\bibitem{Y.Lee2014}
Y. Lee, Y. Hwang, S. B. Cho, and Y.-C. Chung,
Achieving a direct band gap in oxygen functionalized-monolayer scandium carbide by applying an electric field,
Phys. Chem. Chem. Phys. 16 (2014) 26273. 

\bibitem{Y.Lee2015}
Y. Lee, Y. Hwang, and Y.-C. Chung
Achieving type I, II, and III heterojunctions using functionalized MXene,
ACS Appl. Mater. Interfaces 7 (2015) 7163$-$7169.


\bibitem{A.S.Ingason2016}
A. S. Ingason, M. Dahlqvist, J. Rosen,
Magnetic MAX phases from theory and experiments; a review, 
J. Phys.: Condens. Matter 28 (2016) 433003.


\bibitem{G.Gao2016_1} 
G. Gao, G. Ding, J. Li, K. Yao, M. Wu, M. Qian, 
Monolayer MXenes: promising half-metals and spin gapless semiconductors,
Nanoscale 8 (2016) 8986$-$8991.

\bibitem{C.Si2015} 
C. Si, J. Zhou, Z. Sun, 
Half-metallic ferromagnetism and surface functionalization-induced metal-insulator transition in graphene-like two-dimensional Cr$_2$C crystals,
ACS Appl. Mater. Interfaces 7 (2015) 17510$-$17515.

\bibitem{J.He2016} 
J. He, P. Lyu, L. Z. Sun, \'{A}. M. Garc\'{i}a, P. Nachtigall, 
High temperature spin-polarized semiconductivity with zero magnetization in two-dimensional Janus MXenes,
J. Mater. Chem. C 4 (2016) 6500$-$6509.

\bibitem{H.Kumar2017}
H. Kumar, N. C. Frey, L. Dong, B. Anasori, Y. Gogotsi, V. B. Shenoy,
Tunable Magnetism and Transport Properties in Nitride MXenes,
ACS Nano 11 (2017) 7648$-$7655.

\bibitem{L.Dong2017}
L. Dong, H. Kumar, B. Anasori, Y. Gogotsi, V. B. Shenoy,
Rational design of two-dimensional metallic and semiconducting spintronic materials based on ordered double-transition-metal MXenes,
J. Phys. Chem. Lett. 8 (2017) 422$-$428.

\bibitem{W.Chen2017}
W. Chen, H.-F. Li, X. Shi, H. Pan, 
Tension-tailored electronic abd magnetic switching of 2D Ti$_2$NO$_2$,
J. Phys. Chem. C 121 (2017) 25729$-$25735.


\bibitem{S.S.Li2017}   
S.-S. Li, S.-J. Hu, W.-X. Ji, P. Li, K. Zhang, C.-W. Zhang, and S.-S. Yan,
Emergence of ferrimagnetic half-metallicity in two-dimensional MXene Mo$_3$N$_2$F$_2$,
Appl. Phys. Lett. 111 (2017) 202405.



\bibitem{N.C.Frey2018}    
N. C. Frey, H. Kumar, B. Anasori, Y. Gogotsi, V. B. Shenoy,
Tuning noncollinear spin structure and anisotropy in ferromagnetic nitride MXenes,
ACS Nano 12 (2018) 6319$-$6325.


\bibitem{W.Sun2018}   
W. Sun, Y. Xie, and P. R. C. Kent,
Double transition metal MXenes with wide band gaps and novel magnetic properties,
Nanoscale 10 (2018) 11962$-$11968.

\bibitem{J.He2018}
J. He, G. Ding, C. Zhong, S. Li, D. Li, and G. Zhang,
Nanoscale 11 (2019) 356$-$364. 


\bibitem{J.J.Zhang2018}
J.-J. Zhang, L. Lin, Y. Zhang, M. Wu, B. I. Yakobson, and S. Dong,
Type-II Multiferroic Hf$_2$VC$_2$F$_2$ MXene monolayer with high transition temperature,
J. Am. Chem. Soc. 140 (2018) 9768-9773. 




\bibitem{P.Ravindran1999}
P. Ravindran, A. Delin, B. Johansson, O. Eriksson, J. M. Wills, 
Electronic structure, chemical bonding, and optical properties of ferroelectric
and antiferroelectric NaNO$_2$,
Phys. Rev. B 59 (1999) 1776$-$1785.

\bibitem{G.R.Berdiyorov2016}
G. R. Berdiyorov,
Optical properties of functionalized Ti$_3$C$_2$T$_2$ (T= F, O, OH) MXene: First-principles calculations,
AIP Adv. 6 (2016) 055105.

\bibitem{Y.Bai2016}
Y. Bai, K. Zhou, N. Srikanth, J. H. L. Pang, X. He, R. Wang,
Dependence of elastic and optical properties on surface terminated groups in two-dimensional MXene monolayers: a first-principles study,
 RSC Adv. 6 (2016) 35731$-$35739.
 

\bibitem{H.Zhang2016}
H. Zhang, G. Yang, X. Zuo, H. Tang, Q. Yang, G. Li, 
Computational studies on the structural, electronic and optical properties of graphene-like MXenes (M$_2$CT$_2$, M= Ti, Zr, Hf; T= O, F, OH) and their potential applications as visible-light driven photocatalysts,
 J. Mater. Chem. A 4 (2016) 12913$-$12920.



\bibitem{J.Guo2017}
J. Guo, Y. Sun, B. Liu, Q. Zhang, Q. Peng,
Two-dimensional scandium-based carbides (MXene): Band gap modulation and optical properties,
  J. Alloys Compd. 712 (2017) 752$-$759.
  

\bibitem{A.Chandrasekaran2017}
A. Chandrasekaran, A. Mishra, A. K. Singh,
Ferroelectricity, antiferroelectricity, and ultrathin 2D electron/hole gas in multifunctional monolayer MXene,
Nano Lett. 17 (2017) 3290$-$3296.


\bibitem{G.Ying2018}
G. Ying, S. Kota, A. D. Dillon, A. T. Fafarman, M. W.Barsoum,
Conductive transparent V$_2$CT$_x$ (MXene) films,
FlatChem 8 (2018) 25. 

\bibitem{R.Li2017}
R. Li, L. Zhang, L. Shi, and P. Wang,
MXene Ti$_3$C$_2$: an effective 2D light-to-heat conversion material,
ACS Nano 11 (2017) 3752$-$3759.





\bibitem{M.Khazaei2015} 
Khazaei M, Arai M, Sasaki T, Ranjbar A, Liang Y, Yunoki S.
 OH-terminated two-dimensional transition metal carbides and nitrides as ultralow work function materials,
Phys. Rev. B 92 (2015) 075411.

\bibitem{Y.Liu2016}
Y. Liu, H. Xiao, W. A. Goddard III,
Schottky-barrier-free contacts with two-dimensional semiconductors by surface-engineered MXenes, 
J. Am. Chem. Soc. 138 (2016) 15853$-$15856.

\bibitem{M.Khazaei2016_1} 
M. Khazaei, A. Ranjbar, M. Ghorbani-Asl, M. Arai, T. Sasaki, Y. Liang, S. Yunoki,  
Nearly free electron states in MXenes,
Phys. Rev. B 93 (2016) 205125.

\bibitem{Q.Peng2014}
Q. Peng, J. Guo, Q. Zhang, J. Xiang, B. Liu, A. Zhou, R. Liu, Y. Tian,  
Unique lead adsorption behavior of activated hydroxyl group in two-dimensional titanium carbide,
J. Am. Chem. Soc. 136 (2014) 4113$-$4116.

\bibitem{N.Alidoust2014}
N. Alidoust, G. Bian, S.-Y. Xu, R. Sankar, M. Neupane, C. Liu, I. Belopolski, D.-X. Qu, J. D. Denlinger, F.-C. Chou and 
M. Z. Hasan, Observation of monolayer valence band spin-orbit effect and induced quantum well states in MoX$_2$,
Nat. Commu. 5, 4673 (2014). 

\bibitem{M.Otani2010}
M. Otani and S. Okada,
Field-induced free-electron carriers in graphite,
J. Phys. Soc. Jpn 79 (2010) 073701.







\bibitem{M.N.Blonsky2015}
M. N. Blonsky, H. L. Zhuang, A. K. Singh, R. G. Hennig,
Ab initio prediction of piezoelectricity in two-dimensional materials,
ACS Nano 9 (2015) 9885$-$9891.

\bibitem{W.Li2015}
W. Li, J. Li, 
Piezoelectricity in two-dimenional group-III monochalcogenides,
Nano Res. 8 (2015) 3796$-$3802.

\bibitem{C.Sevik2016}
C. Sevik, D. \c{C}akir, O. G\"{u}lseren, F. M. Peeters,
Peculiar piezoelectric properties of soft two-dimensional materials,
J. Phys. Chem. C 120 (2016) 13948$-$13953.

\bibitem{R.Bechmann1958}
R. Behmann,
Elastic and piezoelectric constants of alpha-quartz,
Phys. Rev. 110 (1958) 1060$-$1061.


\bibitem{C.M.Lueng2000}
C. M. Lueng, H. L. W. Chan, C. Surya, C. L. Choy,
Piezoelectric coefficient of aluminum nitride and gallium nitride,
J. Appl. Phys. 88 (2000) 5360$-$5363.





\bibitem{S.Kumar2016}
S. Kumar and U. Schwingenschl\"{o}gl,
Thermoelectric performance of functionalized Sc$_2$C MXenes,
Phys. Rev. B 94 (2016) 035405.

\bibitem{A.N.Gandi2016}
A. N. Gandi, H. N. Alshareef, U. Schwingenschl\"{o}gl,
Thermoelectric performace of the MXenes M$_2$CO$_2$ (M= Ti, Zr, or Hf),
Chem. Mater. 28 (2016) 1647$-$1652.

\bibitem{S. Sarikurt2018}
S. Sarikurt, D. \c{C}akir, M. Ke\c{c}eil, C. Sevik, 
The influence of surface fuctionalization on thermal transport and thermoelectric properties of MXene monolayers, 
Nanoscale 10 (2018) 8859$-$8868.

\bibitem{W.Jeitschko1983}
W. Jeitschko, H. Nowotny, F. Benesovsky,
Kohlenstoffhaltige tern\"{a}re Verbindungen (H-Phase),
Monatsh. Chem. 94 (1963) 672$-$676.

\bibitem{R.Meshkian2015}
R. Meshkian, L. \AA N\"{a}slund, J. Halim, J. Lu, M. W. Barsoum, J. Ros\'{e}n,  
Synthesis of two-dimensional molybdenum carbide, Mo$_2$C, from the gallium based 
atomic laminate Mo$_2$Ga$_2$C,
Scripta Mater. 108 (2015) 147$-$150.

\bibitem{C.Xu2015}
C. Xu, L. Wang, Z. Liu, L. Chen, J. Guo, N. Kang, X.-L. Ma, H.-M. Cheng, W. Ren, 
Large-area high-quality 2D ultrathin Mo$_2$C superconducting crystals,
Nat. Mater. 14 (2015) 1135$-$1141.


\bibitem{D.Geng2017}
D. Geng, X. Zhao, Z. Chen, W. Sun, W. Fu, J. Chen, W. Liu, W. Zhou, K. P. Loh,
Direct synthesis of large-area 2D Mo$_2$C on in situ grown graphene,
Adv. Mater. 29 (2017) 1700072. 

\bibitem{W.Sun2016}
W. Sun, Y. Li, B. Wang, X. Jiang, M. I. Katsnelson, P. Korzhavyi, O. Eriksson, and I. D. Marco,
A new 2D monolayer BiXene,M$_2$C (M=Mo,Tc,Os),
Nanoscale 8 (2016) 15753. 



\bibitem{J.Bardeen1957}
J. Bardeen, L. N. Cooper, J. R. Schrieffer, 
Theory of superconductivity,
Phys. Rev. 108 (1957) 1175$-$1204.

\bibitem{J.Lei2017}
J. Lei, A. Kutana, B. I. Yakobson, 
Predicting stable phase monolayer Mo$_2$C (MXene), 
a superconductor with chemically-tunable critical temperature,
J. Mater. Chem. C 5 (2017) 3438$-$3444.
 
\bibitem{J.J.Zhang2017} 
J.-J. Zhang, S. Dong,
Superconductivity of monolayer Mo$_2$C: the key role of functional groups,
J. Chem. Phys. 146 (2017) 034705.


\bibitem{Y.D.Fu2017}
 Y.-D. Fu,  B. Wang,  Y. Teng,  X.-S. Zhu,  X.-X. Feng,  M.-F. Yan, P. Korzhavyide, and W. Sun,
 The role of group III, IV elements in Nb$_4$AC$_3$ MAX phases (A = Al, Si, Ga, Ge) and the unusual anisotropic behavior of the electronic and optical properties,
 Phys. Chem. Chem. Phys.19 (2017) 15471$-$15483. 





\bibitem{C.Ling2016} 
C. Ling, L. Shi, Y. Ouyang, J. Wang,
  Searching for highly active catalysts for hydrogen evolution reaction based on O-terminated MXenes through a simple descriptor,
  Chem. Mat. 28 (2016) 9026$-$9032.
 
\bibitem{G.Gao2016_2}
G. Gao, A. P. O'Mullane, A. Du,
  2D MXenes: A new family of promising catalysts for the hydrogen evolution reaction,
  ACS Catal. 7 (2016) 494$-$500.

\bibitem{N.Chaudhari2017}
N. K. Chaudhari, H. Jin, B. Kim, and D. S. Baek, S. H. Joo, K. Lee,
MXene: an emerging two-dimensional material for future energy conversion and storage applications,
 J. Mater. Chem. A 5 (2017) 24579$-$24579.

\bibitem{J.C.Lei2015}
J.-C. Lei, X. Zhang, Z. Zhou,
  Recent advances in MXene: Preparation, properties, and applications,
  Front. Phys. 10 (2015) 276$-$286.

\bibitem{Z.Seh2016}
Z. W. Seh, K. D. Fredrickson, B. Anasori, J. Kibsgaard, A. L. Strickler, M. R. Lukatskaya, Y. Gogotsi, T. F. Jaramillo, A. Vojvodic,
  Two-dimensional molybdenum carbide (MXene) as an efficient electrocatalyst for hydrogen evolution,
  ACS Energy Lett. 1 (2016) 589$-$594.

\bibitem{X.An2018}
X. An, W. Wang, J. Wang, H. Duan, J. Shi, X. Yu, Xuelian,
The synergetic effects of Ti$_3$C$_2$ MXene and Pt as co-catalysts for highly efficient photocatalytic hydrogen evolution over g-C$_3$N$_4$,
Phys. Chem. Chem. Phys. 20 (2018) 11405$-$11411.

\bibitem{Y.Sun2018}
Y. Sun, D. Jin, Y. Sun, X. Meng, Y. Gao, Y. Dall'Agnese, G. Chen, X.-F.  Wang,
  g-C$_3$N$_4$/Ti$_3$C$_2$T$_x$ (MXenes) composite with oxidized surface groups for efficient photocatalytic hydrogen evolution,
  J. Mater. Chem. A 6 (2018) 9124$-$9131.



\bibitem{C.Cheng2018}
C. Cheng, X. Zhang, M. Wang, S. Wang, Z. Yang,
  Single Pd atomic catalyst on Mo$_2$CO$_2$ monolayer (MXene): unusual activity for CO oxidation by trimolecular Eley--Rideal mechanism,
  Phys. Chem. Chem. Phys. 20 (2018) 3504$-$3513.


\bibitem{Z.Guo2016}
Z. Guo, J. Zhou, L. Zhu, Z. Sun,
  MXene: a promising photocatalyst for water splitting,
 J. Mater. Chem. A 4 (2016) 11446$-$11452.

\bibitem{A.Shahzad2018}
A. Shahzad, K. Rasool, M. Nawaz, W. Miran, J. Jang, M. Moztahida, K. A. Mahmoud, D. S. Lee,
 Heterostructural TiO$_2$/Ti$_3$C$_2$T$_x$ (MXene) for photocatalytic degradation of antiepileptic drug carbamazepine,
  Chem. Eng. J. 349 (2018) 748$-$755.

\bibitem{J.Low2018}
J. Low, L. Zhang, T. Tong, B. Shen, J. Yu, 
 TiO$_2$/MXene Ti$_3$C$_2$ composite with excellent photocatalytic CO$_2$ reduction activity,
  J. Catal. 361 (2018) 255$-$266.

\bibitem{J.Ran2017}  
J. Ran, G. Gao, F.-T. Li, T.-Y. Ma, A. Du, S.-Z. Qiao,
Ti$_3$C$_2$ MXene co-catalyst on metal sulfide photo-absorbers for enhanced visible-light photocatalytic hydrogen production,
 Nat. Commu. 8 (2017) 13907.
 
 
 \bibitem{M.Naguib2013_2}
 M. Naguib, J. Halim, J. Lu, K. M. Cook, L. Hultman, Y. Gogotsi, M. W. Barsoum,
 New two-dimensional niobium and vanadium carbides as promising materials for Li-Ion batteries,
 J. Am. Chem. Soc. 135 (2013) 15966$-$15969.
 

 
\bibitem{A.VahidMohammadi2017}
A. VahidMohammadi, A. Hadjikhani, S. Shahbazmohamadi, M. Baidaghi,
 Two-dimensional vanadium carbide (MXene) as a high-capacity cathode material for rechargeable aluminum batteries,
 ACS Nano 11 (2017) 11135$-$11144.
 
 \bibitem{Q.Tang2012}
Q. Tang, Z. Zhou, P. Shen, 
 Are MXenes promising Anode materials for Li ion batteries? 
 Computational studies on electronic properties and Li storage capability of Ti$_3$C$_2$ and 
 Ti$_3$C$_2$X$_2$ (X = F, OH) monolayer, 
 J. Am. Chem. Soc. 134 (2012) 16909$-$16916.
 
\bibitem{O.Mashtalir2013}
O. Mashtalir, M. Naguib, V. N. Mochalin, Y. Dall'Agnese, M. Heon, M. W. Barsoum, Y. Gogotsi, 
 Intercalation and delamination of layered carbides and carbonitrides, 
 Nature commun. 4 (2013) 1716.
 

\bibitem{D.Er2014}
D. Er, J. Li, M. Naguib, Y. Gogotsi, V. B. Shenoy, 
Ti$_3$C$_2$ MXene as high capacity electrode material for metal (Li, Na, K, Ca) ion batteries, 
ACS Appl. Mater. Interfaces 6 (2014) 11173$-$11179.

\bibitem{Q.Wan2018}
Q. Wan, S. Li, J.-B. Liu, 
First-principle study of Li-ion storage of functionalized Ti$_2$C monolayer with vacancies, 
ACS Appl. Mater. Interfaces 10 (2018) 6369$-$6377.


\bibitem{Y.Aierken2018}
Y. Aierken, C. Sevik, O. Glseren, F.M. Peeters, D. \c{C}akir, 
MXenes/graphene heterostructure for Li battery applications: a first principles study, 
J. Mater. Chem. A. 6 (2018) 2337$-$2345.

 \bibitem{X.Liu2018}
X. Liu, X. Shao, F. Li, M. Zhao, 
 Anchoring effects of S-terminated Ti2C MXene for lithium-sulfur batteries: A first-principles study, 
 Appl. Surf. Sci. 455 (2018) 522$-$526.
 
\bibitem{D.Rao2017}
D. Rao, L. Zhang, Y. Wang, Z. Meng, X. Qian, J. Liu, X. Shen, G. Qiao, R. Lu, 
 Mechanism of the improved performance of lithium sulfur batteries with MXene-based additives, 
 J. Phys. Chem. C. 121 (2017) 11047.
 
 \bibitem{Y.Xie2014_2}
Y. Xie, Y.Dall'Agnese, M. Naguib, Y. Gogotsi, M. W. Barsoum, H. L. Zhuang, L. Zhuang, P. R. C. Kent, 
 Prediction and characterization of MXene nanosheet anodes for non-Lithium ion batteries, 
 ACS Nano 8 (2014) 9606$-$9615.

\bibitem{Y.X.Yu2016}
Y.-X. Yu, 
Prediction of mobility, enhanced storage capacity, and volume change during sodiation on interlayer-expanded functionalized 
Ti$_3$C$_2$ MXene anode materials for sodium-ion batteries, 
J. Phys. Chem. C. 120 (2016) 5288$-$5296.

\bibitem{X.Chen2016}
X. Chen, Z. Kong, N. Li, X, Zhao, C. Sun, 
Proposing prospects of Ti$_3$CN transition metal carbides (MXenes) as anode in Li-ion battery: a DFT study, 
Phys. Chem. Chem. Phys. 18 (2016) 32937$-$32943.
 
\bibitem{D.Wang2017}
D. Wang, Y. Gao, Y. Liu, D. Jin, Y. Gogotsi, X. Meng, F. Du, G. Chen, Y. Wei, 
 First-principles calculations of Ti$_2$N and Ti$_2$NT$_2$ (T = O, F, OH) monolayers as potential anode materials for Lithium-ion batteries and beyond, 
 J. Phys. Chem. C 121 (2017) 13025$-$13034.
 
\bibitem{J.Hu2014}
J. Hu, B. Xu, C. Ouyang, S.A. Yang, Y. Yao, 
Investigations on V$_2$C and V$_2$CX$_2$ (X = F, OH) 
 monolayer as a promising anode material for Li ion batteries from first-principles calculations, 
 J. Phys. Chem. C. 118 (2014) 24274$-$24281.
  
 
\bibitem{L.Bai2016}
L. Bai, H. Yin, X. Zhang, 
 Energy storage performance of V$_{n+1}$C$_n$ monolayer as electrode material studied by first-principles calculations, 
 RSC Adv. 6 (2016) 54999$-$55006.
 
\bibitem{D.Sun2016}
D. Sun, Q. Hu, J. Chen, X. Zhang, L. Wang, Q. Wu, A. Zhou, 
 Structural transformation of MXene (V$_2$C, Cr$_2$C and Ta$_2$C) with O groups during lithiation: 
 A first-principles investigation, 
 ACS Appl. Mater. Interfaces 8 (2016) 74.
 
\bibitem{J.Zhu2015}
J. Zhu, A. Chroneos, U. Schwingenschl�gl, 
Nb-based MXenes for Li-ion battery applications, 
 Phys. Status Solidi RRL 9 (2015) 726$-$729.
 
\bibitem{L.Bai2018}
L. Bai, H. Yin, L. Wu, X. Zhang, 
 First-principle study of the Nb$_{n+1}$C$_n$T$_2$ systems as electrode materials for supercapacitors, 
 Comput. Mater. Sci. 143 (2018) 225$-$231.
 
\bibitem{F.Li2016}
F. Li, C. R. Cabrera, J. Wang, Z. Chen, 
 A Cr$_2$CO$_2$ monolayer as a promising cathode for lithium and non-lithium ion batteries: a computational exploration, 
 RSC Adv. 6 (2016) 81591$-$81596.
 
\bibitem{Z.Zou2018}
X. Zou, G. Li, Q. Wang, D. Tang, B. Wu, X. Wang, 
 Energy storage properties of selectively functionalized Cr-group MXenes, 
 Comput. Mater. Sci. 150 (2018) 236$-$243.
 
\bibitem{Q.Sun2016}
Q. Sun, Y. Dai, Y. Ma, T. Jing, W. Wei, B. Huang, 
 Ab initio prediction and characterization of Mo$_2$C monolayer as anodes for lithium-ion and sodium-ion batteries, 
 J. Phys. Chem. Lett. 7 (2016) 937$-$943.
 
 

 \bibitem{I.Persson2018_2}
I. Persson, J. Halim, H. Lind, T. W. Hansen, J. B. Wagner,
L.-\AA~N\"{a}slund, V. Darakchieva, J. Palisaitis, J. Rosen, and
P. O.~\AA. Persson, 
2D transition metal carbides (MXenes) for carbon capture, 
Adv. Mater. (2018), 1805472. DOI: 10.1002/adma.201805472.

 \bibitem{E.Lee2017}
 E. Lee, A. VahidMohammadi, B. C. Prorok, Y. S. Yoon, M. Beidaghi, and D. J. Kim, 
 Room temperature gas sensing of two-dimensional titanium
carbide (MXene),
 ACS Appl. Mater. Interfaces 9 (2017) 37184.

 
\bibitem{X.F.Yu2015}
X.-F. Yu, Y.-C. Li, J.-B. Cheng, Z.-B. Liu, Q.-Z. Li, W.-Z. Li, X. Yang, B. Xiao, 
Monolayer Ti$_2$CO$_2$: A promising candidate for NH$_3$ sensor or capture with high sensitivity and selectivity, 
ACS Appl. Mater. Interfaces 7 (2015) 13707$-$13713.

\bibitem{A.Junkaew2018}
A. Junkaew, R. Arr\'{o}yave, 
Enhancement of the selectivity of MXenes (M$_2$C, M = Ti, V, Nb, Mo) 
via oxygen-functionalization: Promising materials for gas-sensing and separation, 
Phys. Chem. Chem. Phys. 20 (2018) 6073$-$6082.

\bibitem{S.Ma2017}
S. Ma, D. Yuan, Z. Jiao, T. Wang, X. Dai, 
Monolayer Sc$_2$CO$_2$: A promising candidate as a SO$_2$ gas sensor or capture, 
J. Phys. Chem. C. 121 (2017) 24077$-$24084.
 
 
 
\bibitem{Q.Hu2013}
 Q. Hu, D. Sun, Q. Wu, H. Wang, L. Wang, B. Liu, A. Zhou, J. He, 
MXene: A new family of promising hydrogen storage medium, 
J. Phys. Chem. A. 117 (2013) 14253$-$14260.

\bibitem{Q.Hu2014}
Q. Hu, H. Wang, Q. Wu, X. Ye, A. Zhou, D. Sun, L. Wang, B. Liu, J. He, 
Two-dimensional Sc$_2$C: A reversible and high-capacity hydrogen storage material predicted by first-principles calculations, 
Int. J. Hydrogen Energy 39 (2014) 10606$-$10612.

\bibitem{A.Yadav2016}
A. Yadav, A. Dashora, N. Patel, A. Miotello, M. Press, D.C. Kothari, 
Study of 2D MXene Cr$_2$C materials for hydrogen storage using density functional theory, 
Appl. Surf. Sci. 389 (2016) 88$-$95.



\bibitem{X.Sang2016}
X. Sang, Y. Xie, M.-W. Lin, M. Alhabeb, K. L. V. Aken, Y. Gogotsi, P. R. C. Kent, K. Xiao, and R. R. Unocic,
Atomic defects in monolayer titanium carbide (Ti$_3$C$_2$T$_x$) MXene,
ACS Nano 10 (2016) 9193$-$9200.

\bibitem{S.Zhao2015}
S. Zhao, W. Kang, and J. Xue,
MXene nanoribbons, 
J. Mater. Chem. C 3 (2015) 879.  

\bibitem{Q.Xu2018}
Q. Xu, L. Ding, Y. Wen, W. Yang, H. Zhou,
X. Chen, J. Street, A. Zhou, W.-J. Ong, and N. Li,
High photoluminescence quantum yield of 18.7$\%$ by
using nitrogen-doped Ti3C2 MXene quantum dots,
J. Mater. Chem. C 6 (2018) 6360$-$6369.

\bibitem{A.Kakekhani2016}
A. Kakekhani and S. Isamil-Beigi,
Polarization-driven catalysis via ferroelectric oxide surfaces,
Phys. Chem. Chem. Phys. 18 (2016) 19676$-$19695.  

 
 
 
\end{thebibliography}
\end{document}